\documentclass[aps,prb,twocolumn,superscriptaddress]{revtex4}

\usepackage{lipsum}
\usepackage{graphicx}
\usepackage{amsmath, amsfonts, amssymb, amsbsy}
\usepackage[usenames,dvipsnames]{xcolor}
\usepackage{marvosym}
\usepackage{sidecap}
\usepackage{wrapfig}
\usepackage[utf8]{inputenc}
\usepackage[russian,english]{babel}
\usepackage{hyperref}
\allowdisplaybreaks

\newcommand{\bise} {Bi$_2$Se$_3$}
\newcommand{\bite} {Bi$_2$Te$_3$}
\newcommand{\Seo} {Se$_{\rm out}$}
\newcommand{\Sei} {Se$_{\rm in}$}
\newcommand{\Teo} {Te$_{\rm out}$}
\newcommand{\Tei} {Te$_{\rm in}$}
\newcommand{\grad} {\boldsymbol\nabla}

\newcommand{\drm} {\mathrm{d}}
\newcommand{\kv} {{\bf k}}

\newcommand{ \parS }[1]{ \left[ #1 \right]}

\begin{document}

\title{Tight-binding theory of NMR shifts in topological insulators Bi$_2$Se$_3$ and Bi$_2$Te$_3$}

\author{Samuel Boutin}
\email{Samuel.Boutin@USherbrooke.ca}
\affiliation{D\'{e}partement de Physique and Regroupement Qu\'{e}b\'{e}cois sur les Matériaux de Pointe, Universit\'{e} de Sherbrooke, Sherbrooke, Qu\'{e}bec, Canada J1K 2R1}
\author{Jorge Ram\'irez-Ruiz}
\affiliation{D\'{e}partement de Physique and Regroupement Qu\'{e}b\'{e}cois sur les Matériaux de Pointe, Universit\'{e} de Sherbrooke, Sherbrooke, Qu\'{e}bec, Canada J1K 2R1}
\affiliation{Instituto de F\'isica, Universidad Nacional Aut\'onoma de M\'exico, Apartado Postal 20-364, 01000 M\'exico D.F., México}
\author{Ion Garate}
\affiliation{D\'{e}partement de Physique and Regroupement Qu\'{e}b\'{e}cois sur les Matériaux de Pointe, Universit\'{e} de Sherbrooke, Sherbrooke, Qu\'{e}bec, Canada J1K 2R1}

\date{\today}

\begin{abstract}
Motivated by recent nuclear magnetic resonance (NMR) experiments, we present a microscopic $sp^3$ tight-binding model calculation of the NMR shifts in bulk \bise\, and \bite.
We compute the contact, dipolar, orbital and core polarization contributions to the carrier-density-dependent part of the NMR shifts in $^{209}$Bi, $^{125}$Te and $^{77}$Se. The spin-orbit coupling and the layered crystal structure result in a contact Knight shift with strong uniaxial anisotropy.  
Likewise, because of spin-orbit coupling, dipolar interactions make a significant contribution to the isotropic part of the NMR shift. The contact interaction dominates the isotropic Knight shift in $^{209}$Bi NMR, even though the electronic states at the Fermi level have a rather weak $s$-orbital character.
In contrast, the contribution from the contact hyperfine interaction to the NMR shift of $^{77}$Se and $^{125}$Te is weak compared to the dipolar and orbital shifts therein.
In all cases, the orbital shift is at least comparable to the contact and dipolar shifts, while the shift due to core polarization is subdominant (except for Te nuclei located at the inversion centers).
By artificially  varying the strength of spin-orbit coupling, we evaluate the evolution of the NMR shift across a band inversion but find no clear signature of the topological transition.
\end{abstract}


\maketitle


\section{Introduction}

Nuclear magnetic resonance (NMR) constitutes a powerful experimental technique to study a wide array of chemical and electronic properties of materials.\cite{nmr}
For decades, NMR experiments have provided important information on ordered phases of matter such as superconductivity and magnetism.

In the last five years, NMR has been applied to the study of topological phases of matter,\cite{franz2013} which are characterized by topological invariants that result in peculiar surface states.
{\em A priori}, NMR is not the best tool to characterize topological phases of matter.
On one hand, NMR is a local probe, while topological invariants are nonlocal.
On the other hand, NMR is a bulk probe, while the main manifestations of topological invariants occur at the boundaries/surfaces of the material.

Yet, the aforementioned difficulties have not dissuaded numerous experimental attempts\cite{young2012, nisson2013, nisson2014, mukhopadhyay2015, taylor2012, koumoulis2013, podorozhkin2015, georgieva2016, nowak2014, shi2014, mcfarlane2014, koumolis2014, koumolis2015, koumolis2015b, chasapis2015} aiming at direct or indirect signatures of topological invariants in NMR.
Most of these experiments have focused on \bise\, and \bite, which are model topological insulators.
To mention but a few, Koumoulis {\em et al.}\cite{koumoulis2013, koumolis2015} and Podorozhkin {\em et al.}\cite{podorozhkin2015} searched for NMR signals of topological surface states.
Nowak {\em et al.}\cite{nowak2014} measured a positive $^{209}$Bi NMR shift for topologically trivial YPdBi, and a strongly negative shift for topologically nontrivial YPtBi.
They considered this difference to be a ``possibly universal fingerprint'' of a change in the strong topological invariant.  
This idea was echoed by Shi {\em et al.}\cite{shi2014}, who proposed that it is possible to detect electronic band inversions through NMR shifts.
Koumoulis {\em et al.}\cite{koumolis2015b} used NMR to address the band inversion in certain topological crystalline insulators.
Mukhopadhyay {\em et al.}\cite{mukhopadhyay2015} reported a large isotropic Knight shift for $^{209}$Bi in $n-$doped \bise, which they attributed to the contact hyperfine coupling, with an estimated contact hyperfine coupling ``comparable to or higher than the values determined for GaAs electron systems exhibiting pure $s$-like wave functions.''
These authors suggested that the orbital shift is negligible.
In contrast, MacFarlane {\em et al.}\cite{mcfarlane2014} noted that the diamagnetic orbital response of the free carriers may be important in topological materials where diamagnetism is particularly strong.
More recently, Georgieva {\em et al.}\cite{georgieva2016} ascribed the large magnetic-field-independent NMR linewidth observed by various groups to an unusually strong indirect nuclear coupling mediated by bulk electrons.
It was argued that the reason behind this observation could be an unusually strong Bloembergen-Rowland coupling between nuclear spins in \bise.

On the theoretical front, aside from early work on lead salts\cite{senturia1970, hewes1973, sapoval1973, leloup1973, tripathi1982} predating the discovery of topological phases of matter, there has been a lack of detailed calculations concerning NMR shifts in doped topological crystals.
To our knowledge, the only theoretical activities along this direction consist of just a few continuum model calculations, mainly concerned with the topological surface state contribution to the Knight shift.\cite{zocher2013, schlottmann2014}
Both of these works consider the contact interaction between the nuclei and the itinerant electrons as the only source of the Knight shift.
In addition, neither Ref.~[\onlinecite{zocher2013}] nor Ref.~[\onlinecite{schlottmann2014}] make an attempt to incorporate the variation of the shift from one nucleus to another, or to calculate the hyperfine coupling (instead leaving it as a phenomenological parameter). 

Partly due to the lack of theoretical guidance, some of the aforementioned experiments appear to offer conflicting responses to a number of basic questions.
Is the contact hyperfine interaction dominant or not in the observed Knight shifts?
Is the orbital shift significant or negligible?
Does core polarization play a major or a secondary role?
What is the influence of spin-orbit interactions and band inversions in NMR?
To quote Ref.~[\onlinecite{mukhopadhyay2015}], ``a better understanding [...] would require new theoretical studies''.
The objective of this paper is to present a theory that addresses these questions by predicting the carrier-density-dependent portion of the contact, dipolar, core polarization and orbital contributions to the NMR shift. 

The rest of this work is organized as follows.
In Sec.~\ref{sec:TBmodel}, we review the microscopic $sp^3$ tight-binding theory of bulk \bise\, and \bite, which is our method of choice for the calculation of NMR shifts.
Compared to earlier ${\bf k}\cdot{\bf p}$ model calculations in lead salts,\cite{senturia1970, hewes1973, sapoval1973, leloup1973, tripathi1982} the tight-binding approach is more powerful in that it allows to compute the NMR contributions from more electronic bands and from the entire Brillouin zone.
This advantage is especially important in heavily-doped semiconductors, as well as in systems (such as \bise\, and \bite) where the bottom of the conduction band and the top of the valence band are not at the same point in the Brillouin zone.  
Indeed, the popular ${\bf k}\cdot{\bf p}$ model proposed in Ref.~[\onlinecite{liu2010}] would not be a good starting point for the calculation of the NMR shift in $p-$doped \bite, because the top of the valence band is away from the $\Gamma$ point (in fact, it is not located in a high-symmetry point of the Brillouin zone).
The tight-binding theory we present has no such problem.  
In addition, our calculation has the advantage of yielding the $g$ factors and hyperfine couplings without resorting to experiment. 
The main limitation of the $sp^3$ tight-binding theory, which is shared by the ${\bf k}\cdot{\bf p}$ theory,  is that it is conceived to accurately reproduce the electronic bands in the vicinity of the Fermi level, but not those far from it.
Consequently, it can reliably describe the carrier-density-dependent part of the NMR shift, but not the density-independent part that originates from core electrons (also known as the chemical shift).
Density-functional theory (DFT) calculations, commonly used for chemical shifts,\cite{wien2k} do not suffer from this limitation.
However, a DFT calculation of NMR shifts in conducting and strongly spin-orbit coupled crystals remains a computationally difficult task,\cite{abinitio}  not undertaken thus far.
 In view of this, the tight-binding approach strikes a useful compromise which allows us to predict the dependence of NMR shifts on the carrier density.
This dependence is experimentally accessible\cite{hewes1973, nisson2013} by measuring the variation of the resonance field with carrier concentration and subtracting the reference field, i.e. the field at which resonance would occur in absence of carriers.

In Sec.~\ref{sec:knight}, we compute the contact, dipolar and core polarization contributions to the Knight shift, which emerge from a magnetic-field-induced spin polarization of the itinerant carriers. Contrary to what is occasionally assumed, we find that the contribution from the contact interaction to the Knight shift is strongly anisotropic, and that the dipolar interaction does contribute to the {\em isotropic} part of the Knight shift. These two observations are direct consequences of the strong spin-orbit interactions in \bise\, and \bite. We demonstrate that the $^{209}$Bi NMR shift in $n-$doped \bise\, is dominated by the contact interaction; however, for other shifts, the dipolar and orbital parts are often considerable and at times dominant. With the exception of the $^{125}$Te NMR shift,  core polarization is not a major actor. In comparing \bise\, with \bite, we find that the hyperfine coupling for Te nuclei located at the inversion centers is much larger than the hyperfine coupling for Se nuclei located at the same place. As a byproduct of our calculation, we evaluate the contact hyperfine fields for Bi, Te and Se, as well as the $g$ factors for \bise\, (at the $\Gamma$ point), and find a rather good agreement with experiment.

In Sec.~\ref{sec:orbital}, we evaluate the contribution of the valence electrons to the orbital shift, which emerges from magnetic-field-induced orbital currents. The density-dependent part of this shift is comparable to that of the Knight shift. As a result, contact, dipolar, and orbital mechanisms can all be comparably important for the NMR shifts in \bise\, and \bite.

Section~\ref{sec:discussion} summarizes the main results and draws the conclusions. Besides, we discuss the effect of a band inversion in the  carrier-density-dependent part of the NMR shifts, and show that the latter do not contain generic signatures of topological invariants. The technical aspects of the theory are relegated to Appendices~\ref{app:derivationPT}, \ref{app:dipolar}, and \ref{app:orbital}.

\section{Tight-Binding Model}\label{sec:TBmodel}

The crystal structure of \bise\, consists of an ABC stacking of monoatomic triangular lattices normal to the $c$-axis.
These layers are grouped into quintuple layers (QL) of strongly bounded planes, while neighboring QL interact mainly through van der Waals forces. Each QL contains two equivalent ``outer'' Se planes (\Seo), two equivalent Bi planes, and another ``inner'' Se plane (\Sei) located at the center of inversion.\cite{liu2010}
Due to the ABC stacking, the primitive rhombohedral unit cell spans three QL and contains five atoms: two \Seo, one \Sei\, and two Bi. An identical crystal structure applies to \bite, upon replacing Se by Te. 
For all atoms, the valence electrons are in $p$-type orbitals. 

In this work, we adopt a $sp^3$ tight-binding description of the single-electron Hamiltonian with spin-orbit interactions,
\begin{equation}
\label{eq:h0}
{\cal H}=\frac{{\bf p}^2}{2 m}+V({\bf r}) + \frac{\hbar}{4 m^2 c^2} (\grad V \times{\bf p})\cdot{\boldsymbol\sigma},
\end{equation}
where $V({\bf r})=V({\bf r}+{\bf R})$ is the lattice potential, ${\bf R}$ are the Bravais vectors giving the positions of the unit cells, $m$ is the bare electron mass,  ${\bf p}=-i\hbar\grad$ is the canonical momentum, $c$ is the speed of light in vacuum and ${\boldsymbol\sigma}$ is a vector of Pauli matrices denoting the spin degree of freedom.
The electronic velocity operator, which plays a central role in the theory of NMR shifts,  is given by ${\bf v}=(i/\hbar) [ {\cal H}, {\bf r}]
 = {\bf p}/m+(\hbar/4 m c^2) {\boldsymbol\sigma}\times\grad V$.

In the tight-binding description of the electronic structure, each lattice site is ascribed a localized electronic state $|{\bf R} j \mu \sigma\rangle=|{\bf R} j \mu\rangle |\sigma\rangle$, 
$j$ labels the five atoms inside the primitive unit cell, $\mu= \{ s, p_x, p_y, p_z\}$ denotes the atomic orbitals considered in the $sp^3$ model,\cite{kobayashi2011} and $\sigma$ is the spin index.
The states $|{\bf R} j \mu \sigma\rangle $ are L\"owdin orbitals, obeying
$\langle {\bf R} j \mu \sigma | {\bf R}' j' \mu' \sigma'\rangle = \delta_{{\bf R}{\bf R}'} \delta_{j j'} \delta_{\mu\mu'} \delta_{\sigma\sigma'}$.

At first thought, one may be tempted to approximate $|{\bf R} j \mu\rangle$ by free hydrogen-like atomic orbitals with atomic numbers $Z_j$ that correspond to Bi, Se or Te. 
However, this approximation ignores screening effects and thus overestimates the probability of finding an electron at the nucleus by roughly an order of magnitude.
More so,  such an error propagates onto the calculation of NMR shifts and gives, for example, effective hyperfine fields that are $\sim 10$ times larger than the values known from experiment. 
Therefore, we calculate the atomic orbitals' wave functions using the DFT implementation of the first-principles quantum chemistry package Psi4.\cite{psi4}
Our calculation incorporates intra-atomic electron-electron interactions as well as scalar relativistic effects.
As we show below, this approach gives effective hyperfine fields in good agreement with experiment.
Although the calculated orbitals centered in different atoms are not completely orthogonal, the overlaps $\int \drm^3 r \langle{\bf r}|{\bf R} j \mu\rangle \langle {\bf R}' j' \mu' | {\bf r}\rangle$  for $j\neq j'$ are small enough that neglecting them does not result in a significant error.

\begin{figure}[t]
\centering
\includegraphics[width=0.4\textwidth]{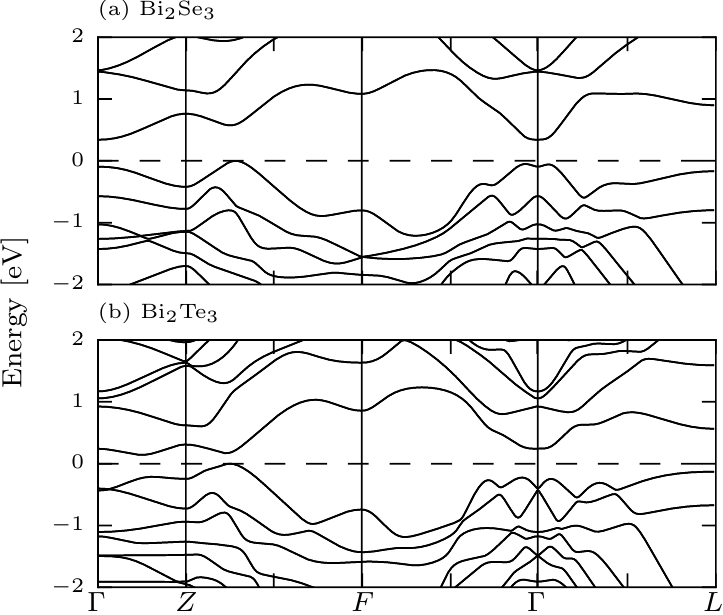}
\caption{Bulk tight-binding bands for \bise\, and \bite.}
\label{fig:bands}
\end{figure}

In the basis of L\"owdin orbitals, the Hamiltonian from Eq.~(\ref{eq:h0}) can be recasted in the second quantized form as
\begin{equation}
\label{eq:Hr}
{\cal H}=\sum_{{\bf R} j \mu \sigma} \sum_{{\bf R}' j' \mu' \sigma'} {\cal H}_{j \mu \sigma; j' \mu' \sigma'} ({\bf R}, {\bf R}') c^\dagger_{{\bf R} j \mu \sigma} c_{{\bf R}' j' \mu' \sigma'},
\end{equation}
where $ c^\dagger_{{\bf R} j \mu \sigma}$ is an operator that creates an electron in state $|{\bf R} j \mu \sigma\rangle$, and ${\cal H}_{j \mu \sigma; j' \mu' \sigma'} ({\bf R}, {\bf R}') = \langle {\bf R} j \mu \sigma | {\cal H} | {\bf R}' j' \mu' \sigma'\rangle$.
In \bise\, and \bite, the numerical values of $\langle {\bf R} j \mu \sigma | {\cal H} | {\bf R}' j' \mu' \sigma'\rangle$ have been tabulated for up to third nearest neighboring sites by fitting to results from DFT.\cite{kobayashi2011}
Spin-orbit interactions, crucial in these materials,  are incorporated through onsite terms.
Fourier transforming Eq.~(\ref{eq:Hr}), we have
\begin{equation}
{\cal H} = \sum_{\bf k} \sum_{j \mu \sigma} \sum_{j' \mu' \sigma'} {\cal H}_{j \mu \sigma; j' \mu' \sigma'} ({\bf k}) c^\dagger_{{\bf k} j \mu \sigma} c_{{\bf k} j' \mu' \sigma'},
\end{equation}
where ${\bf k}$ is the crystal momentum (within the first Brillouin zone), $c^\dagger_{{\bf k} j \mu \sigma}$ is an operator that creates an electron in a Bloch spinor
\begin{equation}
|\psi_{{\bf k} j \mu \sigma}\rangle = \frac{1}{\sqrt{N}} \sum_{\bf R} e^{i {\bf k}\cdot({\bf R}+{\bf t}_j)} |{\bf R} j \mu \sigma\rangle,
\end{equation}
${\bf t}_j$ is the position of a given atom in the unit cell (so that ${\bf R}+{\bf t}_j$ is its actual position in the lattice), and $N$ is the number of unit cells in the crystal.
The eigenstates and eigenvalues of ${\cal H}$ are $|\psi_{{\bf k} n}\rangle$ and $E_{{\bf k} n}$, respectively, where $n$ denotes the band index. 
In particular, $|\psi_{{\bf k} n}\rangle$ are Bloch spinors that obey $\langle {\bf r}+{\bf R}|\psi_{{\bf k} n}\rangle = \exp(i{\bf k}\cdot{\bf R})\langle{\bf r}|\psi_{{\bf k} n}\rangle$ and  $\langle\psi_{{\bf k} n} | \psi_{{\bf k}' n'}\rangle = \delta_{{\bf k} {\bf k}'}\delta_{n n'}$. 
The set of states $\{|\psi_{{\bf k} j \mu \sigma}\rangle\}$ defined in Eq~(4) form an orthonormal basis.
As such, we may write
\begin{equation}
\label{eq:exp}
|\psi_{{\bf k} n}\rangle = \sum_{j \mu \sigma} C_{{\bf k} n;{\bf k} j \mu \sigma} |\psi_{{\bf k} j \mu \sigma}\rangle,
\end{equation}
where $C_{{\bf k} n;{\bf k} j \mu \sigma}=\langle \psi_{{\bf k} j \mu \sigma} | \psi_{{\bf k} n} \rangle$.
The matrix elements of the Hamiltonian ${\cal H}({\bf k})$ are 
\begin{align}
\begin{split}
&{\cal H}_{j \mu \sigma; j' \mu' \sigma'}({\bf k}) = \langle \psi_{{\bf k} j \mu \sigma} | {\cal H} | \psi_{{\bf k} j' \mu' \sigma'}\rangle\\
&=\frac{1}{N}\sum_{{\bf R} {\bf R}'} e^{i {\bf k}\cdot({\bf R}+{\bf t}_j-{\bf R}'-{\bf t}_{j'})}{\cal H}_{j \mu \sigma; j' \mu' \sigma'} ({\bf R}, {\bf R}').
\end{split}
\end{align}
To avoid confusion, we remark that ${\cal H}({\bf k})\neq e^{-i {\bf k}\cdot{\bf r}} {\cal H} e^{ i {\bf k}\cdot{\bf r}}$.
A numerical diagonalization of ${\cal H}({\bf k})$ yields the eigenvalues $E_{{\bf k} n}$ and the coefficients $C_{{\bf k} n;{\bf k} j \mu \sigma}$
 ($n=1,...,40$).
Figure~\ref{fig:bands} displays $E_{{\bf k} n}$ along the high symmetry directions in the first Brillouin zone\cite{kobayashi2011}. 
Each energy level is doubly degenerate due to the combined time-reversal symmetry and spatial inversion symmetry.

In the remainder of this paper, we make use of $E_{{\bf k} n}$ and $|\psi_{{\bf k} n}\rangle$ to estimate the NMR shifts in \bise\, and \bite.
These shifts are induced by a spatially uniform external magnetic field ${\bf B}$ that modifies Eq.~(\ref{eq:h0}) through a Zeeman term $\mu_B {\boldsymbol\sigma}\cdot{\bf B}$ and through the minimal coupling ${\bf p}\to {\bf p}+ e{\bf A}$, where ${\bf A}={\bf B}\times{\bf r}/2$ is the vector potential in the symmetric gauge and $e$ is the absolute value of the bare electronic charge.
This results in   
${\cal H}\to{\cal H}+\delta{\cal H}$, where
\begin{equation}
\label{eq:dH}
\delta{\cal H}= -{\boldsymbol\mu}\cdot{\bf B}+(e^2/8 m) |{\bf B}\times{\bf r}|^2,
\end{equation}
${\boldsymbol\mu}=-\mu_B ({\boldsymbol\sigma}+m {\bf r}\times{\bf v}/\hbar)$ is the magnetic moment of the itinerant electrons and $\mu_B = e\hbar/(2 m)$ is the Bohr magneton. 
Below, we evaluate the NMR shifts to first order in the applied field.
For the Knight shift, this implies that the $\mathcal{O}(B^2)$ term in Eq.~(\ref{eq:dH}) can be ignored.
However, both terms in Eq.~(\ref{eq:dH}) are required for the orbital shift.

%%%%%%%%%%%%%%%%%%%%%%%                                                                                                                                                                                           
%%%%%%%%%%%%%%%%%%%%%%%%%%        

\section{Knight shift}\label{sec:knight}
\subsection{General considerations}
The Knight shift originates from a magnetic-field-induced spin polarization of the itinerant carriers.
NMR shifts due to the orbital motion of electrons will be discussed in the next section.
Let $n_e$ be the bulk electronic density measured from the neutrality point (at zero temperature, $n_e>0$ when the Fermi level intersects the conduction bands, $n_e<0$ when the Fermi level intersects the valence bands, and $n_e=0$ when the Fermi level lies within the gap).   
Throughout this work, we assume that $n_e$ is spatially uniform.
This assumption allows to take into account the average influence of defects and dopants on the NMR shift. 
Spatial inhomogeneities of the carrier concentration in real samples will produce a distribution (linewidth) of Knight shifts around the mean value we calculate.
The details of NMR linewidths are beyond the scope of this work.

The external magnetic field ${\bf B}$ couples to the magnetic moment ${\boldsymbol\mu}$ of the itinerant electrons (cf. Eq.~(\ref{eq:dH})) and induces a spin polarization.
Then, this spin polarization acts on the nuclear spins through contact and dipolar interactions as well as through core polarization effects, thereby producing an extra carrier-density-dependent magnetic field 
\begin{equation}
\label{eq:knight}
\delta{\bf B}^{\rm Knight}({\bf r}_0,n_e) = \sum_\lambda \delta{\bf B}^\lambda({\bf r}_0, n_e),
\end{equation}
where $\lambda \in \{\text{cont, dip, core}\}$ labels the different contributions to the Knight shift, to be defined momentarily.
This additional magnetic field shifts the resonance frequency of a particular nuclear spin $j_0$ located at position ${\bf r}_0={\bf R}_0+{\bf t}_{j_0}$.
Explicitly, the contribution due to the contact interaction is given by\cite{nmr} 
\begin{equation}
\label{eq:Bcont}
\delta {\bf B}^{\rm cont}({\bf r}_0, n_e)=-\frac{2}{3}\mu_0 g_s \mu_B \langle {\bf S} ({\bf r}_0) \rangle, 
\end{equation}
whereas the contribution due to the dipolar interaction reads
\begin{align}
\label{eq:Bdip}
\begin{split}
\delta{\bf B}^{\rm dip}({\bf r}_0, n_e) =&\frac{\mu_0}{4\pi} g_s \mu_B \int \drm^3 r \left[\frac{\langle{\bf S}({\bf r})\rangle}{|{\bf r}-{\bf r}_0|^3}\right.\\
&~\left.-\frac{3 ({\bf r}-{\bf r}_0)}{|{\bf r}-{\bf r}_0|^5} \langle{\bf S}({\bf r})\rangle \cdot ({\bf r}-{\bf r}_0)\right].
\end{split}
\end{align}
Here, $\mu_0$ is the magnetic permeability in vacuum, $g_s=2$ is the {\em bare} electronic $g$-factor,\cite{sapoval1973} and $\langle {\bf S}({\bf r})\rangle$ is the expectation value of the local electronic spin-density operator at position ${\bf r}$,
\begin{equation}
{\bf S}({\bf r})={\boldsymbol \sigma} |{\bf r}\rangle\langle{\bf r}|/2.
\end{equation}
The reason why $g_s=2$ in Eq.~(\ref{eq:Bcont}), regardless of the strength of spin-orbit interactions, is that the contact interaction acts on atomically short lengthscales.
On the other hand, the effective $g$ factor of the band eigenstates (which departs from $2$ in presence of spin-orbit interactions), is reflected in $\langle {\bf S}({\bf r})\rangle$. 
As a result, the Knight shift scales linearly with the effective $g$ factor of the band eigenstates.\cite{sapoval1973} 

The third mechanism for the Knight shift is the core polarization, which originates from a Coulomb-interaction-induced spin polarization of the core $s$ electrons.
Because the valence electrons of \bise\, and \bite\, are predominantly of $p$ type, core polarization effects could in principle be at least as important as the contact interaction.
The effective magnetic field felt by the nucleus due to core polarization can be expressed as\cite{narath} 
\begin{equation}
\label{eq:dbcore}
\delta{\bf B}^{\rm core}({\bf r}_0, n_e)= - B_{\rm eff}^{\rm core}(j_0) \langle{\bf s}_p({\bf r}_0)\rangle,
\end{equation}
where $B_{\rm eff}^{\rm core}(j_0)$ is the effective core hyperfine field per $p$ electron in an open shell and ${\bf s}_p({\bf r}_0)$ is the spin operator projected onto the $p$-orbitals at nucleus $j_0$.
The effective core hyperfine fields (in kOe) are experimentally known\cite{carter1977} to be $-300$, $-150$ and $-50$ for Bi ($6p$), Te ($5p$) and Se ($4p$), respectively; we will adopt these values in our theory.
 Similarly, we note that the contact term of Eq.~(\ref{eq:Bcont}) can be rewritten in a way that is formally identical to Eq.~(\ref{eq:dbcore}):\cite{carter1977}
\begin{equation}
\label{eq:dbcont}
\delta{\bf B}^{\rm cont}({\bf r}_0, n_e)=-B_{\rm eff}^{\rm atom}(j_0) \langle{\bf s}_s({\bf r}_0)\rangle,
\end{equation}
where $B_{\rm eff}^{\rm atom}(j_0) = (2 g_s/3) \mu_0 \mu_B |\langle{\bf r}_0|{\bf R}_0 j_0 s\rangle|^2$ 
is the effective contact hyperfine field per $s$ electron and ${\bf s}_s({\bf r}_0)$ is the spin operator projected onto the $s$-orbitals at nucleus $j_0$. 
We evaluate $B_{\rm eff}^{\rm atom}(j_0)$ through atomic DFT calculations\cite{psi4} and compare the results to available experimental values in Table~\ref{tab:hyperfineField}.
\begin{table}[tb]
\caption{Comparison between the atomic hyperfine fields $B_{\rm eff}^{\rm atom}$ (cf. Eq.~\eqref{eq:dbcont}) calculated in this work and the reference experimental values from Ref.~[\onlinecite{carter1977}].
}\label{tab:hyperfineField}
\begin{ruledtabular}
\begin{tabular}{ccc}
$B_{\rm eff}^{\rm atom}$ (kOe)  & Calculated & Experimental value\\ \hline
Bi & 49,100 &  49,000 \\ \hline
Se & 12,580 &   - \\ \hline
Te & 20,850 &  17,200 
\end{tabular}
\end{ruledtabular}
\end{table}

The Knight shift depends on $n_e$ because $\langle{\bf S}({\bf r})\rangle$ and $\langle {\bf s}_p({\bf r}_0)\rangle$ depend on $n_e$.
Given that $\langle{\bf S}({\bf r})\rangle$ and $\langle {\bf s}_p({\bf r}_0)\rangle$ are linear in the external magnetic field (for weak fields), it is customary to rewrite Eq.~(\ref{eq:knight}) as 
\begin{equation}
\label{eq:dk0}
\delta B_i^\lambda({\bf r}_0,n_e)\equiv \sum_j K^{\rm \lambda}_{i j}(j_0, n_e) B_j \text{   ,  ($i, j = x, y, z$)}.
\end{equation}
Here, $K^{\rm cont}(j_0, n_e)$, $K^{\rm dip}(j_0, n_e)$ and $K^{\rm core}(j_0, n_e)$ are the dimensionless Knight shift tensors for the contact, dipolar and core polarization at carrier density $n_e$ and at a nucleus $j_0$, respectively.
Due to the axial symmetry of the problem, $K^\lambda_{i j}\propto \delta_{i j}$ and $K^\lambda_{xx}=K^\lambda_{yy}\neq K^\lambda_{zz}$.

As discussed in the Introduction,  we focus on the carrier-density-dependent part of the shifts, i.e. 
\begin{equation}
\label{eq:dk}
\Delta K_{i i}^{\rm \lambda} (j_0, n_e)\equiv K^{\rm \lambda}_{i i} (j_0, n_e) - K^{\rm \lambda}_{i i}(j_0, 0).
\end{equation}
Then, the zero of the Knight shift corresponds to the situation where there are no free carriers in the system (i.e. at a temperature low compared to the bulk gap and with the Fermi energy inside the bulk bandgap). 
Equation~(\ref{eq:dk}) can be experimentally accessed by measuring the NMR shifts as a function of carrier concentration (cf. Sec.~\ref{sec:discussion}).
\cite{yesinowski}
The chemical shift, independent of the carrier density, is cancelled out in Eq.~(\ref{eq:dk}).
Since the chemical shift is independent of temperature as well, it can also be separated out through temperature-dependent  measurements of the NMR shift.

\subsection{Perturbation theory expression for the itinerant spin density}

Assuming a pristine crystal and recalling that the nuclear resonance frequency is negligible compared to electronic energy scales, the average local spin density to first order in ${\bf B}$ is given by standard perturbation theory (cf. App.~\ref{app:derivationPT}),
\begin{equation}
\label{eq:S}
\langle {\bf S}({\bf r}) \rangle \simeq 
\sum_{{\bf k}{\bf k}' n n'} \langle \psi_{{\bf k} n}| {\bf S}({\bf r})| \psi_{{\bf k}' n'}\rangle\langle \psi_{{\bf k}' n'} | \delta{\cal H} | \psi_{{\bf k} n}\rangle f_{n n'}({\bf k},{\bf k}'),
\end{equation}
where $\delta{\cal H}$ includes only the first term in the right hand side of Eq.~(\ref{eq:dH}), ${\bf k}$ and ${\bf k}'$ are within the first Brillouin zone,
\begin{equation}
\label{eq:fnn}
f_{n n'}({\bf k},{\bf k}')\equiv (f_{{\bf k} n}-f_{{\bf k}' n'})/(E_{{\bf k} n}- E_{{\bf k}' n'})
\end{equation}
and $f_{{\bf k} n}$ is the Fermi-Dirac occupation number for a Bloch state with energy $E_{{\bf k} n}$.
The calculation of $\langle {\bf s}_p ({\bf r}_0)\rangle$, relevant for the core polarization, follows an identical derivation.
Equation~(\ref{eq:S}) contains a non-interacting expression for the local spin susceptibility. 
The exchange enhancement of the spin susceptibility can be ignored, because the static dielectric constant of \bise\, and \bite\, is large ($\sim 100$).

In Eq.~(\ref{eq:S}), it is illustrative to separate the sums over the band indices into intraband ($E_{{\bf k} n} = E_{{\bf k} n'}$) and interband ($E_{{\bf k} n} \neq E_{{\bf k} n'}$) parts, which are associated with Pauli and Van Vleck susceptibilities, respectively.
When ${\bf k}={\bf k}'$, intraband transitions give a contribution to the NMR shift that scales as the density of states at the Fermi energy,\cite{dfdE} while the contribution from interband transitions depends relatively weakly on the carrier density and remains nonzero even when the Fermi energy lies within the bulk gap.
In experiments,\cite{nisson2013} it is customary to identify the Knight shift with the part of the NMR shift that depends on carrier density (i.e. largely the intraband part).
The density-independent part of the shift, frequently believed to have an orbital origin,  is customarily lumped together with the chemical shift.
However, in strongly spin-orbit coupled materials such as \bise\, and \bite, the Van Vleck term makes a large density-independent contribution to the itinerant {\em spin} density.

The numerical evaluation of Eq.~(\ref{eq:S}) requires calculating the matrix elements of $\delta{\cal H}$ to first order in $B$.
This task is somewhat delicate because $\delta{\cal H}$ contains the unbounded operator ${\bf r}$. 
There exist various approaches to treat this issue;\cite{r} here, we follow the method of Ref.~[\onlinecite{blount1962}], according to which the matrix elements of the position operator in the Bloch states can be written as $\langle\psi_{{\bf k}' n'}|{\bf r}|\psi_{{\bf k} n}\rangle = -i \grad_{\bf k} \langle\psi_{{\bf k}' n'}|\psi_{{\bf k} n}\rangle + i \delta_{{\bf k}{\bf k}'}\langle u_{{\bf k} n'}|\grad_{\bf k} u_{{\bf k}n}\rangle_{\rm cell}$, where $|u_{{\bf k} n}\rangle = \sqrt{N} e^{-i {\bf k}\cdot{\bf r}} |\psi_{{\bf k} n}\rangle$ is the lattice-periodic part of the Bloch spinor, and the subscript ``cell'' means that the spatial integral is carried out within a unit cell.
Using this, and saving the details of the derivation for Appendix~\ref{app:derivationPT}, Eq.~(\ref{eq:S}) becomes
\begin{equation}
\label{eq:ssum}
\langle {\bf S}({\bf r})\rangle = \sum_{j=1}^4\langle {\bf S}({\bf r})\rangle_j,
\end{equation}
where
\begin{widetext}
\begin{align}
\label{eq:S2}
\begin{split}
\langle S_i({\bf r})\rangle_1 &= \mu_B \sum_{{\bf k}}\sum_{n n'} \langle\psi_{{\bf k} n} | S_i({\bf r}) | \psi_{{\bf k} n'}\rangle \langle \psi_{{\bf k} n'} |{\boldsymbol\sigma}\cdot{\bf B} |\psi_{{\bf k} n}\rangle f_{n n'}({\bf k},{\bf k})\\
\langle S_i({\bf r})\rangle_2 &= m \mu_B {\rm Im} \sum_{{\bf k}}\sum_{ n n'} \langle\psi_{{\bf k} n} |S_i({\bf r}) | \psi_{{\bf k} n'}\rangle  f_{n n'}({\bf k},{\bf k})
\sum_{E_{n''}\neq E_{n'}}\frac{(\langle\psi_{{\bf k} n'}| {\bf v} |\psi_{{\bf k} n''}\rangle\times\langle\psi_{{\bf k} n''}|{\bf v}|\psi_{{\bf k} n}\rangle)\cdot{\bf B}}{E_{{\bf k} n'}-E_{{\bf k} n''}}\\
\langle S_i({\bf r})\rangle_3 &=\frac{m\mu_B}{\hbar} {\rm Im} \sum_{{\bf k}}\sum_{ n n'} \langle\psi_{{\bf k} n}|S_i({\bf r})|\psi_{{\bf k} n'}\rangle  f_{n n'}({\bf k},{\bf k}) \sum_{E_{n''}=E_{n'}} (\langle\grad_{\bf k} u_{{\bf k} n'}|u_{{\bf k} n''}\rangle_{\rm cell} \times \langle\psi_{{\bf k} n''}|{\bf v}|\psi_{{\bf k} n}\rangle)\cdot{\bf B}\\
\langle S_i({\bf r})\rangle_4 &=-\frac{m\mu_B}{\hbar}{\rm Im}\sum_{{\bf k}{\bf k}'}\sum_{n n'} \delta_{{\bf k} {\bf k}'} \left[\grad_{{\bf k}} \times \left(\langle\psi_{{\bf k} n} | S_i ({\bf r}) | \psi_{{\bf k}' n'}\rangle f_{n n'}({\bf k},{\bf k}')\langle \psi_{{\bf k} n'} | {\bf v} | \psi_{{\bf k} n}\rangle\right)\right]\cdot{\bf B}.
\end{split}
\end{align}
\end{widetext}
The terms $\langle{\bf S}({\bf r})\rangle_1$, $\langle{\bf S}({\bf r})\rangle_2$ and $\langle{\bf S}({\bf r})\rangle_3$ only involve transitions that are diagonal in the crystal momentum, unlike $\langle{\bf S}({\bf r})\rangle_4$.
In the expression for $\langle {\bf S}({\bf r})\rangle_4$, one must take the derivative with respect to ${\bf k}$ {\em before} applying Kronecker's $\delta_{{\bf k} {\bf k}'}$.
The expression $\langle{\bf S}({\bf r})\rangle$ is invariant under gauge transformations of the form $|\psi_{{\bf k} n}\rangle \to U_{{\bf k} n} |\psi_{{\bf k} n}\rangle$, where $U_{{\bf k} n}$ is any unitary matrix, differentiable with respect to ${\bf k}$, acting on the twofold degenerate subspace of band $n$ at momentum ${\bf k}$.
It is worth noting that $\langle{\bf S}({\bf r})\rangle_1$ and $\langle{\bf S}({\bf r})\rangle_2$ are separately gauge-invariant, while only the sum of $\langle{\bf S}({\bf r})\rangle_3$ and $\langle{\bf S}({\bf r})\rangle_4$ is gauge-invariant. 

In absence of spin-orbit interactions, we find $\langle{\bf S}({\bf r})\rangle_2=\langle{\bf S}({\bf r})\rangle_3=\langle{\bf S}({\bf r})\rangle_4=0$, and the interband part of $\langle{\bf S}({\bf r})\rangle_1$ vanishes as well.
What is left in this case is the textbook expression for the Knight shift in simple metals,\cite{nmr} which scales like the Pauli spin susceptibility with an electronic $g$ factor of $2$.
Nevertheless, since \bise\, and \bite\, are strongly spin-orbit coupled narrow-gap semiconductors, it is insufficient to consider solely the intraband part of $\langle{\bf S}({\bf r})\rangle_1$.
For one thing, in presence of spin-orbit interactions, the effective $g$ factor of the electronic bands departs strongly from $2$. 
The intraband part of $\langle{\bf S}({\bf r})\rangle_2$ quantifies how such a departure affects the Knight shift. 
Thus, the intraband part of $\langle{\bf S}({\bf r})\rangle_1+\langle{\bf S}({\bf r})\rangle_2$ has been used repeatedly in calculations of the Knight shift in spin-orbit coupled semiconductors (see e.g. Ref.~[\onlinecite{tripathi1982}]).
 
The meaning of $\langle{\bf S}({\bf r})\rangle_3$ and $\langle{\bf S}({\bf r})\rangle_4$ is less intuitive. 
These terms originate from spin-orbit interactions and, in that sense, appear to be related to the ``spin-orbit contribution'' discussed from a different viewpoint in Ref.~[\onlinecite{tripathi1982}].
Insofar as we restrict our attention to the intraband transitions ($E_{{\bf k} n}=E_{{\bf k} n'}$), and insofar as the temperature is low and the Fermi level lies close to a band extremum at ${\bf k}_0$ (a circumstance that is common in self-doped semiconductors such as \bise\, and \bite), then $\langle {\bf S}({\bf r})\rangle_3+\langle {\bf S}({\bf r})\rangle_4$ is small.
This is because $\langle\psi_{{\bf k}_0 n} |{\bf v}|\psi_{{\bf k}_0 n'}\rangle = \delta_{n n'}\partial E_{{\bf k} n}/(\hbar \partial {\bf k}) |_{{\bf k}={\bf k}_0} = 0$ for $E_{{\bf k}_0 n}=E_{{\bf k}_0 n'}$. 
Yet, when a semiconductor is moderately doped, the intraband parts of $\langle{\bf S}({\bf r})\rangle_3$ and $\langle{\bf S}({\bf r})\rangle_4$ can make a sizeable impact. 
Our expression for the intraband part of $\langle {\bf S}({\bf r})\rangle_3$ can be directly connected to early theoretical attempts\cite{yafet1963, beuneu1980} to determine how the $g$ factor changes away from a band extremum.
These theories were nonetheless aware of the lack of gauge-invariance of this term.
This is where the importance of $\langle {\bf S}({\bf r})\rangle_4$ becomes manifest in our theory, as it restores the overall gauge invariance.
We stress that in order to obtain $\langle{\bf S}({\bf r})\rangle_4$, it is essential to allow for ${\bf k}\neq {\bf k}'$ in Eq.~(\ref{eq:S}).

Equation~(\ref{eq:S2}) is applicable to any single-particle electronic structure, be it based on ${\bf k}\cdot{\bf p}$, tight-binding or DFT methods.
In the tight-binding formalism that concerns us here, the various matrix elements appearing in Eq.~(\ref{eq:S2}) are straightforward to compute.
For example, 
\begin{equation}
\langle \psi_{{\bf k} n'}|{\boldsymbol \sigma}|\psi_{{\bf k} n}\rangle =\sum_{j \mu\sigma\sigma'}  C_{{\bf k} n';{\bf k} j \mu\sigma}^* C_{{\bf k}n;{\bf k} j \mu\sigma'}\langle\sigma|{\boldsymbol \sigma}|\sigma'\rangle,
\end{equation}
where we have used $\langle {\bf R} j \mu \sigma | {\boldsymbol\sigma} |{\bf R}' j' \mu' \sigma'\rangle= \delta_{{\bf R} {\bf R}'} \delta_{j j'} \delta_{\mu\mu'} \langle \sigma|{\boldsymbol \sigma} |\sigma'\rangle$.
Likewise, the matrix elements of the velocity operator are
\begin{align}
\label{eq:v}
\begin{split}
&\langle\psi_{{\bf k} n_1} | {\bf v}|\psi_{{\bf k} n_2}\rangle = \sum_{j\mu\sigma}\sum_{j'\mu'\sigma'} C_{{\bf k}n_1;{\bf k} j \mu\sigma}^* C_{{\bf k}n_2;{\bf k} j' \mu'\sigma'}\\ 
&\left(\grad_{\bf k} {\cal H}_{j \mu \sigma, j'\mu'\sigma'}({\bf k})/\hbar+ i\omega_{{\bf k}n_1; {\bf k}n_2} {\bf d}_{j\mu\sigma, j'\mu'\sigma'} \right),
\end{split}
\end{align}
where $\omega_{{\bf k}n_1; {\bf k} n_2}=(E_{{\bf k} n_1}-E_{{\bf k} n_2})/\hbar$ and ${\bf d}_{j\mu\sigma, j'\mu'\sigma'} = \delta_{\sigma\sigma'}\delta_{j j'} \langle{\bf 0} j \mu |{\bf r}|{\bf 0} j \mu'\rangle$ is the intra-atomic dipole matrix element.\cite{cruz1999} In the derivation of Eq.~(\ref{eq:v}), we have used ${\bf v}=i[{\cal H}, {\bf r}]/\hbar$ and   
${\bf r}|{\bf R} j \mu \sigma\rangle=({\bf R}+{\bf t}_j)|{\bf R} j \mu\sigma\rangle + \sum_{\mu'} {\bf d}_{\mu\mu'} |{\bf R} j \mu' \sigma\rangle$.
Incorporating ${\bf d}$ into the theory ensures nonzero matrix elements of ${\bf v}$ in the limit where the atoms are infinitely far appart from one another. 
However, we find that ${\bf d}$ has a minor quantitative impact in the Knight shift of \bise\, and \bite; hence, it will be ignored hereafter.
Finally, the matrix elements of the local spin density operator read
\begin{align}
\label{eq:shf}
\begin{split}
&\langle \psi_{{\bf k} n}| {\bf S}({\bf r})| \psi_{{\bf k} n'}\rangle \simeq\frac{1}{2 N}\sum_{\sigma\sigma'}\langle \sigma|{\boldsymbol \sigma}|\sigma'\rangle\sum_{{\bf R} j}\sum_{\mu \mu'}\\
&~~~\langle {\bf R} j \mu |{\bf r} \rangle\langle {\bf r} | {\bf R} j \mu'\rangle
C_{{\bf k}n;{\bf k} j \mu \sigma}^* C_{{\bf k}n';{\bf k} j \mu' \sigma'}, 
\end{split}
\end{align}
where we have neglected the overlaps between wave functions localized at different atoms.
Moreover, because $|{\bf R} j \mu\rangle$ are localized, the main contribution to the sum over ${\bf R}$ and $j$ in Eq.~(\ref{eq:shf}) comes from the nucleus nearest to the point ${\bf r}$.
Additional details concerning the numerical evaluation of Eq.~(\ref{eq:S2}) can be found in Appendices~\ref{app:derivationPT} and~\ref{app:dipolar}.

%%%%%%%%%%%%%%%%%%%%%%%                                                                                                                                                                                         
\begin{figure*}[t]
\includegraphics[width=0.4\textwidth]{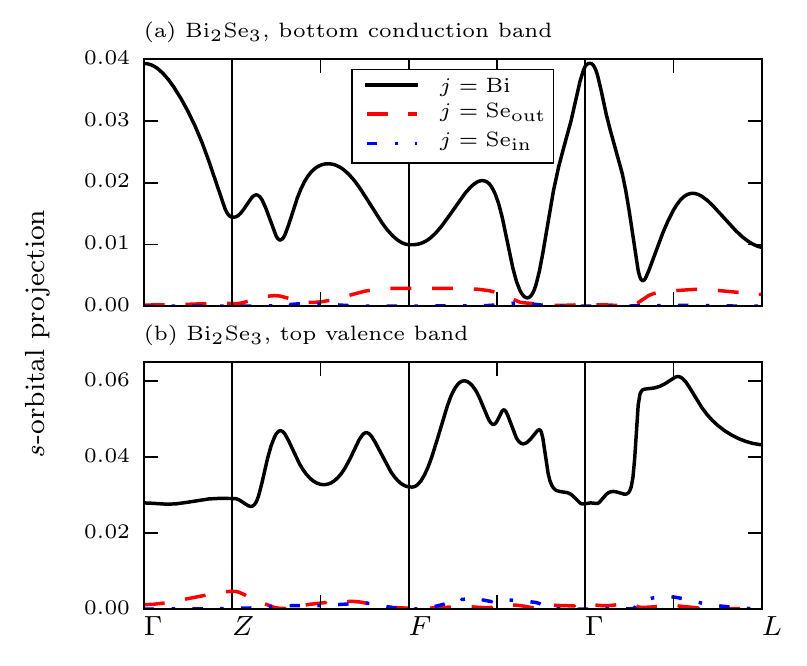}
\includegraphics[width=0.4\textwidth]{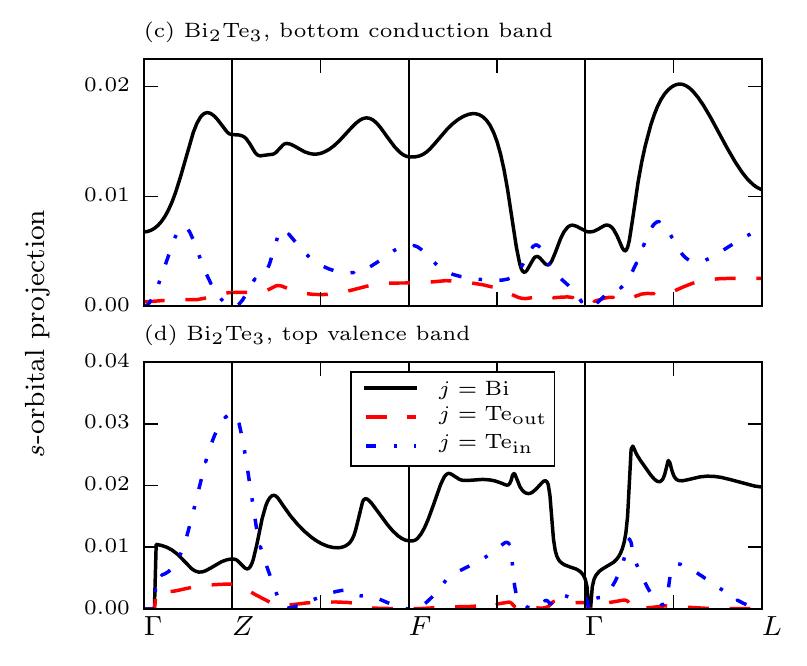}
\caption{The $s-$orbital projection of the tight-binding wave functions, $\sum_\sigma |C_{{\bf k} n; j s \sigma}|^2$, as a function of ${\bf k}$ for the different atoms $j$ in the unit cell, where $n$ is either the lowest conduction band or the highest valence band.
For \bise\, (panels (a) and (b)), our results at $\Gamma$ agree with those of Ref.~[\onlinecite{pertsova}].
Parity is a good quantum number at the time-reversal invariant momenta ($\Gamma, Z, F, L$).
In those points, the full wave function of an odd-parity band has a vanishing projection onto the centers of inversion (\Sei\, and \Tei\, for \bise\, and \bite, respectively).
Recalling that $\sum_{j \mu \sigma} |C_{{\bf k} n; j \mu \sigma}|^2 = 1$, this figure confirms that the low-energy bands have a weak $s-$type character (of the order of 1\% for Bi, and markedly less for the other atoms).
Relative to \bise, the lowest conduction band and the highest valence band of \bite\, have a significantly higher $s-$orbital character on the \Tei\, site (panels c and d).
This, together with the fact that $\langle {\bf r}_0 | {\bf R}_0 {\rm Te}\,, s\rangle > \langle {\bf r}_0 | {\bf R}_0 {\rm Se}\,, s\rangle$, explains why the contact Knight shift is larger in $^{125}{\rm Te}$ than in $^{77}{\rm Se}$.
}
\label{fig:hf}
\end{figure*}
%%%%%%%%%%%%%%%%%%%%%%%%%%        

\subsection{Estimates of the $g$ factors}

As a partial reality check of our theory, we determine the electronic $g$ factors at band extrema, $g_{n \alpha} ({\bf k}_0)$, where $n$ is the band label and $\alpha=1,2,3$ denotes the principal values.
In order to do so, let $|\psi_{{\bf k}_0 n +}\rangle$ and $|\psi_{{\bf k}_0 n -}\rangle$ be the two Bloch states of energy $E_{{\bf k}_0 n}$ in absence of a field, related to one another by the product of time-reversal and spatial inversion operations.
Following Ref.~[\onlinecite{yafet1963}], $[g_{n\alpha} ({\bf k}_0)]^2$ are given by the eigenvalues of the real symmetric tensor ${\cal G}_{i j}=\sum_l G_{i l} G_{l j}$, where 
{\begin{equation}
	\begin{split}
	-\mu_B G_{i x}&=2 {\rm Re}\langle\psi_{{\bf k}_0 n +} | \mu_i |\psi_{{\bf k}_0 n -}\rangle
	\\
	-\mu_B G_{i y}&=-2 {\rm Im}\langle\psi_{{\bf k}_0 n +} | \mu_i |\psi_{{\bf k}_0 n -}\rangle
	\\
	-\mu_B G_{i z}&=2 \langle\psi_{{\bf k}_0 n +} | \mu_i |\psi_{{\bf k}_0 n +}\rangle,
	\end{split}
\end{equation}
and the indices $(i,j,l)$ run over $(x,y,z)$.
Here,  $x$ and $y$ are the crystallographic axes in the plane of the quintuple layers, while $z$ is the axis perpendicular to them.
The matrix elements needed to obtain ${\cal G}$ are computed according to 
\begin{align}
\label{eq:mu}
\begin{split}
&\langle\psi_{{\bf k}_0 n'}|{\boldsymbol\mu}|\psi_{{\bf k}_0 n}\rangle=-\langle\psi_{{\bf k}_0 n'}|{\boldsymbol\sigma}|\psi_{{\bf k}_0 n}\rangle\\
&+i m \sum_{E_{n''}\neq E_n}\frac{\langle\psi_{{\bf k}_0 n'}|{\bf v}|\psi_{{\bf k}_0 n''}\rangle \times\langle\psi_{{\bf k}_0 n''}|{\bf v}|\psi_{{\bf k}_0 n}\rangle}{E_{{\bf k}_0 n}-E_{{\bf k}_0 n''}},
\end{split}
\end{align}
where $E_{{\bf k}_0 n}=E_{{\bf k}_0 n'}$ and the sum is over $n''$.
In absence of spin-orbit interactions, we have verified that $g_{n\alpha}({\bf k})=2$ for all ${\bf k}$.
From the point of view of the Knight shift, the most relevant $g$ factors are those at the bottom of the conduction band (for $n-$doped samples) and at the top of the valence band (for $p-$doped samples).
In \bise, we find $\langle |g_{n 1} ({\bf 0})|, |g_{n 2} ({\bf 0})|, |g_{n 3} ({\bf 0})|\rangle = \langle 16, 16, 20\rangle$  when $n$ is the lowest conduction band, and 
$\langle |g_{n 1} ({\bf 0})|, |g_{n 2} ({\bf 0})|, |g_{n 3} ({\bf 0})|\rangle = \langle 17, 17, 35\rangle$ when $n$ is the highest valence band. 
At ${\bf k}=0$, the principal axes $1,2$ and $3$ coincide with $x,y$ and $z$, respectively; such correspondence no longer applies away from the $\Gamma$ point.
Although the signs of individual $g_{n\alpha}$ can vary depending on convention, the sign of $g_{n 1} g_{n 2}  g_{n 3}$ is invariant and physically measurable.\cite{abragam1970}
In \bise, $sgn[g_{n 1}{\bf 0}) g_{n 2}({\bf 0})  g_{n 3}({\bf 0})]=-1 (+1)$ for the top of the valence band (bottom of the conduction band).\cite{numerics} 
Consequently, as we show below, the Knight shift can change sign if \bise\, transitions from being electron-doped to being hole-doped.

In absolute value, our calculated $g$ factors fall slightly short from the early experimental estimates\cite{kohler1975} in \bise ($\langle 23,23,32\rangle$ for the bottom of the conduction band at the $\Gamma$ point),  but lie within the expected range from recent experiments. \cite{fauque2013}
On the other hand, our results differ significantly from those predicted theoretically by the ${\bf k}\cdot{\bf p}$ method,\cite{liu2010} where a much larger XXZ anisotropy was found in the $g$ factors.
While our work was being finalized, first-principles results for the $g$ factor of \bise\, have appeared in the literature;\cite{fang} the LDA and GGA methods therein give $g$ factors that are in good quantitative agreement with ours.
%%%%%%%%%%%%%%%%%%%%%%%                                                                                                                                                                                           
\begin{figure*}[t]
\includegraphics[width=1\textwidth]{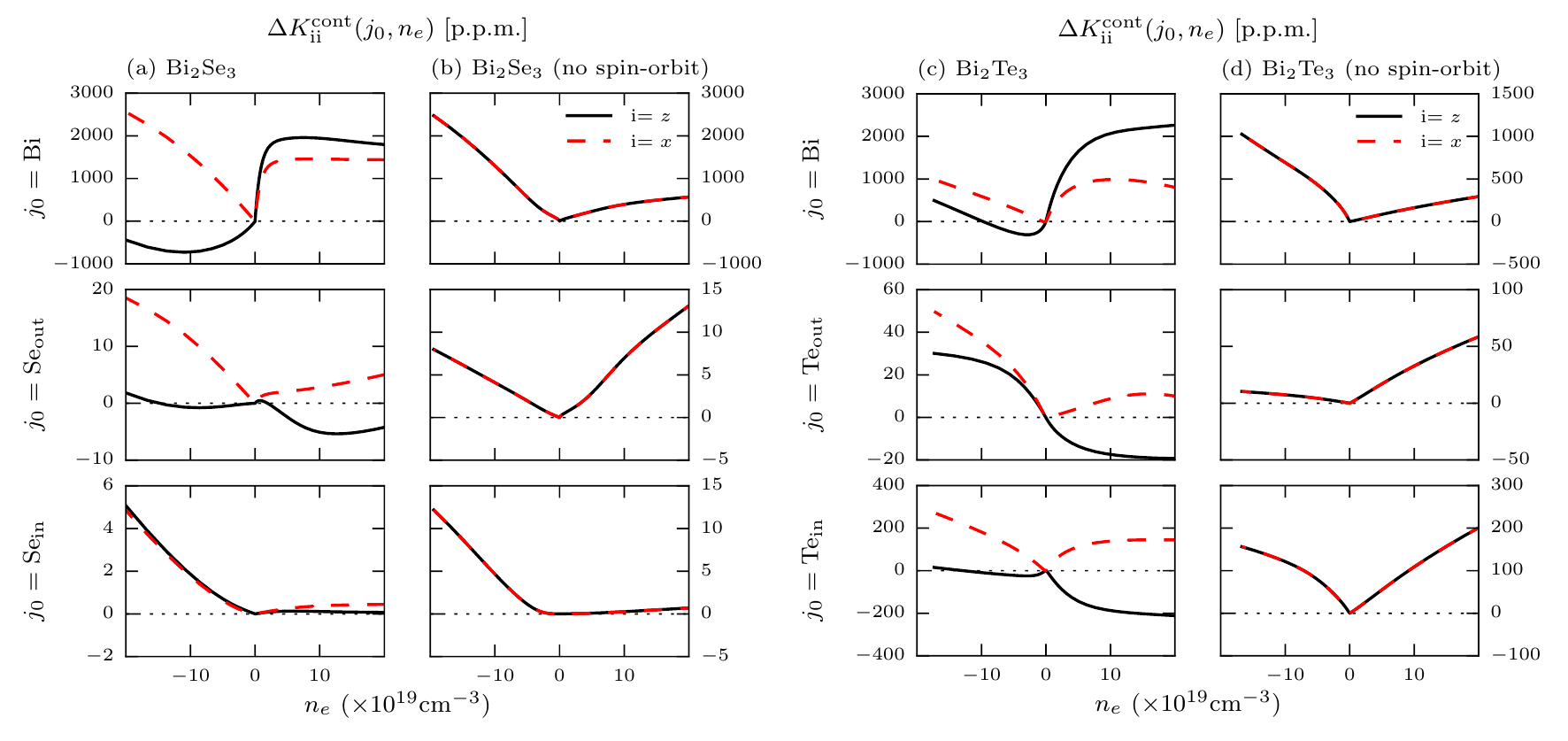}
\caption{Room-temperature contribution of the contact interaction to the Knight shift of \bise\, and \bite, in parts per million (p.p.m.),  as a function of carrier density $n_e$ (cf. Eq.~(\ref{eq:dk})).
For illustration, we include panels where the spin-orbit coupling is turned off by hand, for which $K_{zz}^{\rm cont}=K^{\rm cont}_{xx}$. 
}
\label{fig:kcont}
\end{figure*}
%%%%%%%%%%%%%%%%%%%%%%%%%%        

%%%%%%%%%%%%%%%%%%%%%%%                                                                                                                                                            
\begin{figure*}[t]
\includegraphics[width=1\textwidth]{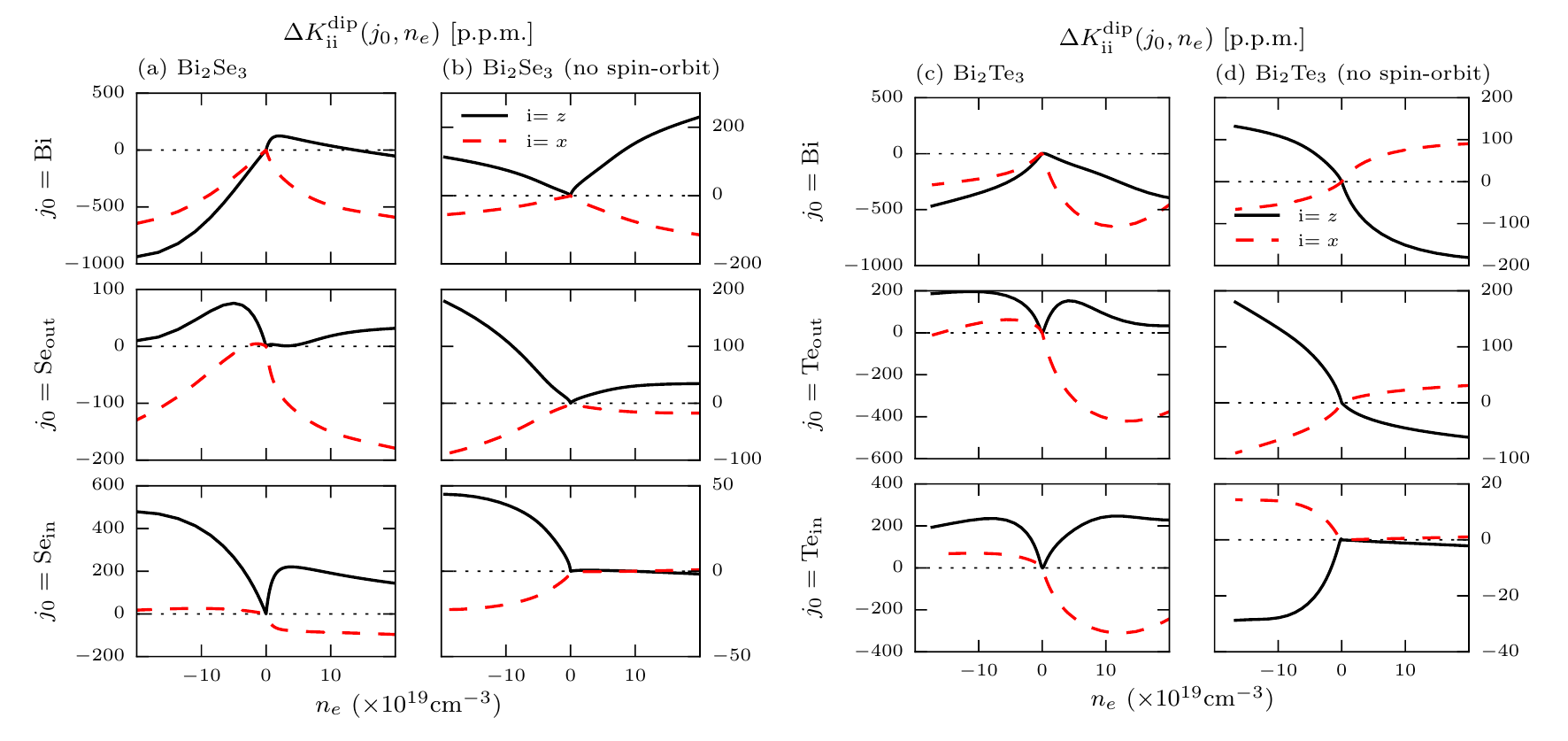}
\caption{Room-temperature contribution of the dipolar interaction to the Knight shift of \bise\, and \bite, in parts per million,  as a function of carrier density $n_e$ (cf. Eq.~(\ref{eq:dk})).
For illustration, we include panels where the spin-orbit coupling is turned off by hand, for which $K_{xx}^{\rm dip}=-K^{\rm dip}_{zz}/2$.}
\label{fig:kdip}
\end{figure*}
%%%%%%%%%%%%%%%%%%%%%%%%%%        

%%%%%%%%%%%%%%%%%%%%%%%                                                                                                                                                            
\begin{figure*}[t]
\includegraphics[width=1\textwidth]{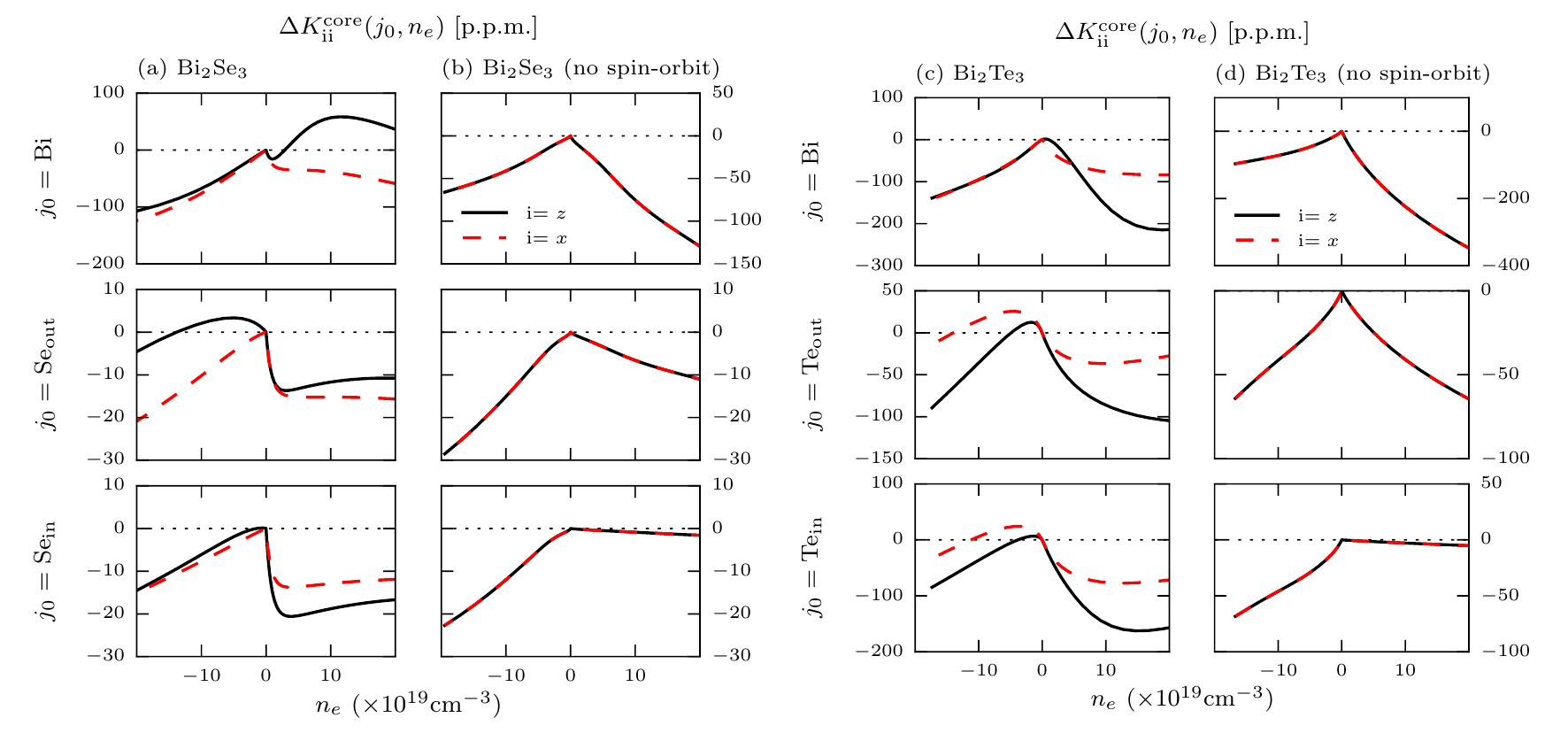}
\caption{Room-temperature contribution of the core polarization to the Knight shift of \bise\, and \bite, in parts per million,  as a function of carrier density $n_e$ (cf. Eq.~(\ref{eq:dk})).
For illustration, we include panels where the spin-orbit coupling is turned off by hand, for which $K_{zz}^{\rm core}=K^{\rm core}_{xx}$.}
\label{fig:kcore}
\end{figure*}
%%%%%%%%%%%%%%%%%%%%%%%%%%        

\subsection{Results}

Let us return to the Knight shift. 
For the contact interaction contribution to the NMR shift (Eq.~(\ref{eq:Bcont})), we are interested in $\langle{\bf S}({\bf r}_0)\rangle$, where ${\bf r}_0={\bf R}_0+{\bf t}_{j_0}$ is the position of a particular nucleus $j_0$.
Then, within the tight-binding approximation, only ${\bf R}={\bf R}_0$ and $j=j_0$ contribute to the sum in Eq.~(\ref{eq:shf}). 
In addition, we set $\mu=\mu'=s$ in Eq.~(\ref{eq:shf})  because $p$ orbitals have vanishing wavefunctions at the nucleus.
Consequently, the contact shift becomes sensitive to factors of the type $|\langle{\bf r}_0|{\bf R}_0 j_0 s\rangle|^2 |C_{{\bf k}n;{\bf k} j_0 s \sigma}|^2$, which provides a quantitative measure of the contact hyperfine coupling associated to the band eigenstate $|\psi_{{\bf k} n}\rangle$ at the nuclear site $j_0$.
There are two aspects that influence this coupling.
First, $\sum_\sigma|C_{{\bf k}n;{\bf k} j_0 s \sigma}|^2$ depends on how ``$s-$like'' $|\psi_{{\bf k} n}\rangle$ is when projected onto the site of atom $j_0$.
Figure~\ref{fig:hf} shows $\sum_\sigma|C_{{\bf k}n;{\bf k} j_0 s \sigma}|^2$ for the highest valence band and lowest conduction band. 
The orbital character of these bands is predominantly of $p$-type both in \bise\, and \bite, which suppresses the contact hyperfine coupling.
Between different atoms, $\sum_\sigma|C_{{\bf k}n;{\bf k} j_0 s \sigma}|^2$ is by far largest for Bi, followed by Te and Se.
Second, $|\langle{\bf r}_0|{\bf R}_0 j_0 s\rangle|^2$ depends on the atomic number and the principal quantum number of the $s$ orbital centered in the atom $j_0$. 
Heavier atoms enhance $|\langle{\bf r}_0|{\bf R}_0 j_0 s\rangle|^2$, which is hence largest for Bi, followed by Te and Se.
Consequently, we anticipate (and corroborate below) that the contact Knight shift is much larger in $^{209}$Bi than in either $^{77}$Se or $^{125}$Te. 

Figures~\ref{fig:kcont}, \ref{fig:kdip} and \ref{fig:kcore} display the Knight shift tensors in \bise\, and \bite, and illustrate two qualitative points that have been either missed or at least not emphasized in recent experiments on topological materials.
First, the contact interaction contribution to the NMR shift has a strong XXZ anisotropy, which originates from combined crystalline anisotropy and spin-orbit interactions. 
This anisotropy is most prominent in Se and Te, where $\Delta K_{xx}^{\rm cont}$ and $\Delta K_{zz}^{\rm cont}$ can even have opposite signs.
Second, the trace of $\Delta K^{\rm dip}$ does {\em not} vanish (i.e. $K_{xx}^{\rm dip}\neq -K_{zz}^{\rm dip}/2$), again as a consequence of spin-orbit interactions.
Accordingly,  dipolar interactions do contribute to the isotropic part of the Knight shift, $(K_{zz}+ 2 K_{xx})/3$.
On a separate note, we remark that the product  $\Delta K^{\rm cont}_{xx} \Delta K^{\rm cont}_{zz}$ changes sign for Bi, \Seo\, and \Teo, but not for \Sei\, and \Tei, when transitioning from electron-doped to hole-doped samples.
This result is a consequence of spin-orbit interactions, in absence of which the signs of $\Delta K^{\rm cont}$ and $\Delta K^{\rm dip}$ are independent of $n_e$.

From a quantitative standpoint, $\Delta K^{\rm cont}({\rm Bi},n_e)$ is positive for $n-$doped samples and is of the order of $\sim 0.2\%$ for $n_e\simeq 5\times 10^{19}{\rm cm}^{-3}$.
Notably, in spite of the dominant $p-$type character of the low-energy states in \bise\, and \bite, $|\Delta K^{\rm cont}({\rm Bi}, n_e)|$ is of the same order as $|\Delta K^{\rm dip} ({\rm Bi},n_e)|$.
In contrast, when it comes to Te and (especially) Se nuclei, the dipolar shift is much larger than the contact shift.
While comparing \bise\, and \bite, the most salient feature is that $|\Delta K^{\rm cont} ({\rm Te}_{\rm in})|\gg|\Delta K^{\rm cont} ({\rm Se}_{\rm in})|$.
The underlying reason for this, explained in Fig.~\ref{fig:hf}, is that the hyperfine coupling for \Tei\, is much larger than the hyperfine coupling for \Sei.
This result is consistent with the experimental findings of Ref.~[\onlinecite{taylor2012}], where it was observed that the isotropic Knight shifts and the spin-lattice relaxation rates are much larger for $^{125}$Te (in \bite) than for $^{77}$Se (in \bise). 
Concerning the core polarization, $\Delta K^{\rm core} ({\rm Bi})$ is at least one order of magnitude smaller than $\Delta K^{\rm cont}({\rm Bi})$. 
The reason for this result may be that $B^{\rm atom}_{\rm eff} \simeq 100 B^{\rm core}_{\rm eff}$ (cf. Table~\ref{tab:hyperfineField} and Eq.~(\ref{eq:dbcore})), while the s-orbital projection at low-energy bands is only $\sim 10$ smaller than the $p-$orbital projection. 
Similarly, the core polarization shift in Se is small compared to the contact and dipolar shifts.
However, core polarization plays a larger role in the case of Te.   

%%%%%%%%%%%%%%%%%%%%%%%                                                                                                                                                            
\begin{figure*}[t]
\includegraphics[width=1\textwidth]{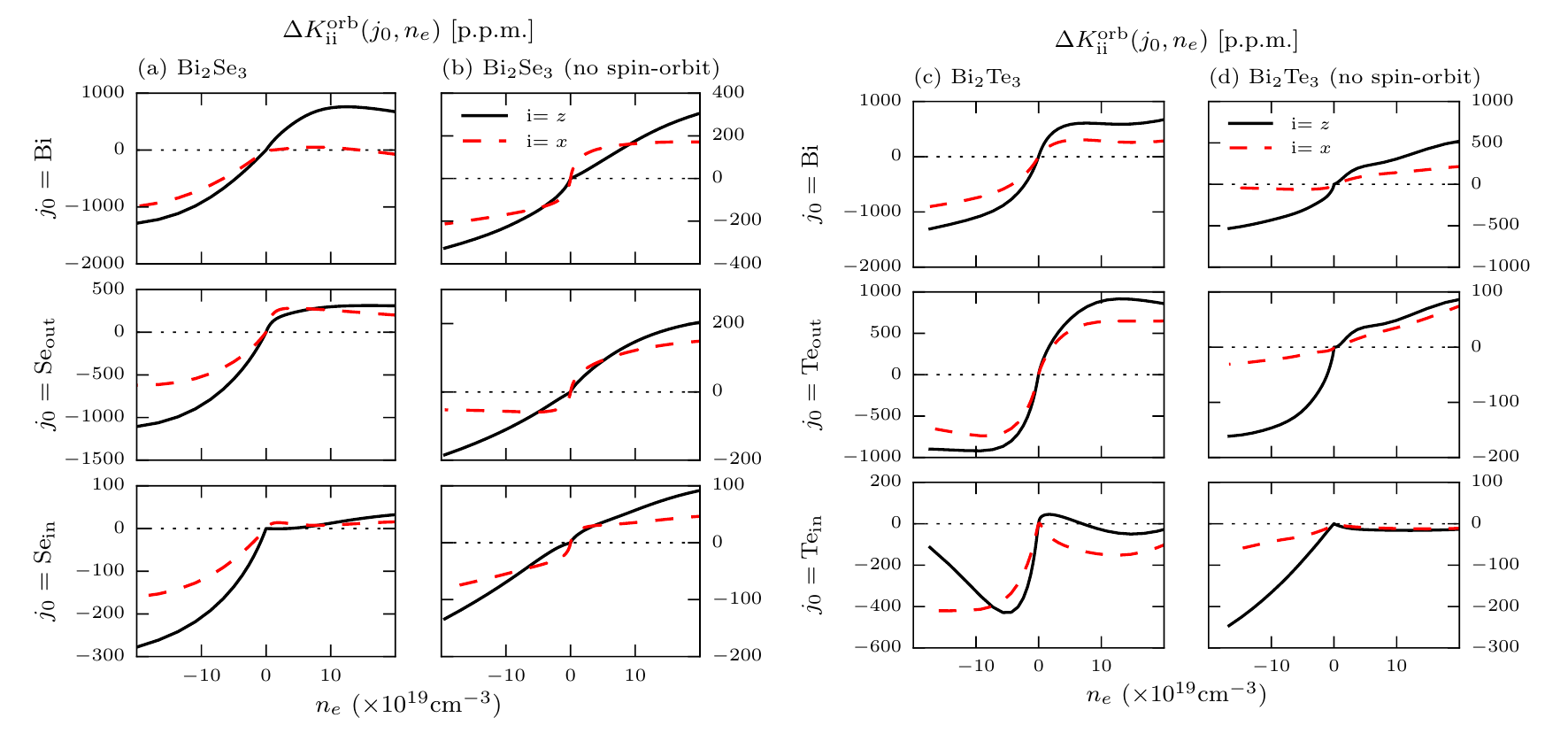}
\caption{Room-temperature orbital shift of \bise\, and \bite, in parts per million,  as a function of carrier density $n_e$ (cf. Eq.~(\ref{eq:dk2})).
For illustration, we include panels where the spin-orbit coupling is turned off by hand.}
\label{fig:korb}
\end{figure*}
%%%%%%%%%%%%%%%%%%%%%%%%%%        

%%%%%%%%%%%%%%%%%%%%%%%                                                       

\section{Orbital shift}\label{sec:orbital}

Thus far we have analyzed the NMR shift due to the coupling between the nuclear spin and the spin density of the itinerant electrons. 
It is well-known\cite{nmr} that there is an additional NMR shift that originates from the orbital currents of the itinerant electrons.
Namely, an applied magnetic field ${\bf B}$ induces an electric current density $\langle{\bf J}({\bf r})\rangle$, which in turn produces a magnetic field 
\begin{equation}
\label{eq:dbo}
\delta {\bf B}^{\rm orb}({\bf r}_0)=\frac{\mu_0}{4\pi}\int \drm^3 r\, \langle{\bf J}({\bf r})\rangle\times\frac{{\bf r}_0-{\bf r}}{|{\bf r}_0-{\bf r}|^3},
\end{equation}
acting on the nuclear spin located at ${\bf r}_0$.
This extra magnetic field constitutes the orbital shift, and the associated dimensionless Knight shift tensor $K^{\rm orb}$ is defined much like in Eq.~(\ref{eq:dk0}),
\begin{equation}
\delta B_i^{\rm orb} ({\bf r}_0, n_e) \equiv \sum_j K^{\rm orb}_{ i j} (j_0, n_e) B_j,
\end{equation}
where $i, j = x, y, z$.
Similarly to the case of the Knight shift, we have $K^{\rm orb}_{i j}\propto \delta_{i j}$ and $K^{\rm orb}_{xx}=K^{\rm orb}_{yy}\neq K^{\rm orb}_{zz}$ due to the axial symmetry of the problem.
% we have 
The objective of this section is to compute 
\begin{equation}
\label{eq:dk2}
\Delta K_{i i}^{\rm orb} (j_0, n_e)\equiv K^{\rm orb}_{i i} (j_0, n_e) - K^{\rm orb}_{i i}(j_0, 0),
\end{equation}
the carrier-density-dependent part of the shift.

To do so, we separate the current density operator ${\bf J}({\bf r})$ into ${\bf J}({\bf r})= {\bf J}_P({\bf r}) + {\bf J}_D ({\bf r})$,
where
\begin{equation}
\label{eq:JP}
{\bf J}_P ({\bf r})=- e\{{\bf v},|{\bf r}\rangle\langle{\bf r}|\}/2
\end{equation}
is the paramagnetic current operator,  $\{ (\cdot), (\cdot) \}$ $\{ \dots, \dots \}$ is an anticommutator, and
\begin{equation}
\label{eq:JD}
{\bf J}_D ({\bf r})=- (e^2/m){\bf A}({\bf r})|{\bf r}\rangle\langle{\bf r}|
\end{equation}
is the diamagnetic current operator.
Only the sum of the two currents is gauge invariant.
Standard linear response in ${\bf A}$ yields (cf. App.~\ref{app:derivationPT})
\begin{align}
\label{eq:j}
	\langle{\bf J}({\bf r})\rangle 
	&= - \frac{ e^2 }{m} {\bf A}({\bf r}) \sum_{{\bf k} n} f_{{\bf k} n} \langle\psi_{{\bf k} n}|{\bf r}\rangle\langle {\bf r}|\psi_{{\bf k} n}\rangle 
	\\&\hspace{-16pt} 
	+\sum_{{\bf k} {\bf k}' n n'} \langle \psi_{{\bf k} n}|{\bf J}_P({\bf r})|\psi_{{\bf k}' n'}\rangle\langle \psi_{{\bf k}'n'}| \delta{\cal H}|\psi_{{\bf k}n}\rangle f_{n n'}({\bf k}, {\bf k}'). \nonumber
\end{align}
In Eq.~(\ref{eq:j}), $\delta{\cal H}$ includes the Zeeman term $\mu_B {\boldsymbol\sigma}\cdot{\bf B}$, which produces orbital currents in presence of spin-orbit interactions.
Although both terms in the right hand side (rhs) of Eq.~(\ref{eq:j}) diverge for $r\to\infty$, their sum converges.\cite{avezac2007}
As a result, for the purposes of our estimates, we assume that the main contribution to the spatial integral in Eq.~(\ref{eq:dbo}) comes from the current density in the vicinity of  ${\bf r}={\bf r}_0$. 
Clearly, the second term on the rhs of Eq.~(\ref{eq:j}) is formally identical to Eq.~(\ref{eq:S}); therefore, we compute it using Eqs.~(\ref{eq:ssum}) and (\ref{eq:S2}), with ${\bf J}_P$ in place of ${\bf S}$.
However, some of the statements made below Eq.~(\ref{eq:S2}), while valid for the local spin density, no longer apply for the orbital current density.
For example, $\langle{\bf J}_P({\bf r})\rangle_3$ and $\langle{\bf J}_P({\bf r})\rangle_4$ are no longer zero in absence of spin-orbit interactions.
Likewise, $\langle {\bf J}({\bf r})\rangle$ depends on the direction of the magnetic field even in absence of spin-orbit coupling, while $\langle{\bf S}({\bf r})\rangle$ does not.
These differences have to do with the fact that the XXZ anisotropy of the lattice is inherited by all orbital observables (like the charge current), while it is communicated to the spin response only in presence of spin-orbit interactions.

In the tight-binding model, we approximate the first term of the right hand side of Eq.~(\ref{eq:j}) with 
\begin{align}
\label{eq:rp}
\begin{split}
&\langle\psi_{{\bf k} n} |{\bf r}\rangle\langle{\bf r}|\psi_{{\bf k}n}\rangle
\simeq \frac{1}{N}\sum_{{\bf R}j}\sum_{\mu\mu'\sigma} \\
&~~~~C^*_{{\bf k} n ;{\bf k} j \mu\sigma} C_{{\bf k} n; {\bf k} j \mu'\sigma}\langle{\bf R} j \mu|{\bf r}\rangle\langle{\bf r}|{\bf R} j \mu'\rangle, 
\end{split}
\end{align}
where the main contribution to the sum over ${\bf R}$ and $j$ comes from the nucleus nearest to the point ${\bf r}$.
Similarly, a simple way to estimate the matrix elements of the paramagnetic current operator is
\begin{equation}
\langle \psi_{{\bf k} n} | {\bf J}_P({\bf r})|\psi_{{\bf k}'n'}\rangle\simeq - \frac{i e \hbar} {2 m} \langle\psi_{{\bf k} n}|{\bf r}\rangle \overleftrightarrow{\nabla} \langle{\bf r}|\psi_{{\bf k}' n'}\rangle,
\end{equation}
where $f\overleftrightarrow{\nabla} g = (\grad f) g - f \grad g$ and we have approximated $\langle{\bf r}|{\bf v} |{\bf R} j \mu \rangle \simeq (-i\hbar/m) \grad \langle{\bf r}|{\bf R} j \mu\rangle$.
This approximation, which neglects the spin-orbit part of the velocity operator,  is qualitatively valid if the most important values of ${\bf r}$ in Eq.~(\ref{eq:dbo}) are close to the nuclear position ${\bf r}_0$, where the electronic wave function is approximately that of an isolated atom.
A more accurate but cumbersome treatment would involve using $\langle{\bf r}|{\bf v} |{\bf R} j \mu \rangle = (i/\hbar)\langle{\bf r}|[{\cal H},{\bf r}] |{\bf R} j \mu \rangle$.  
Additional details concerning the numerical calculation of Eq.~(\ref{eq:j}) may be found in Appendix~\ref{app:orbital}.

Figure~\ref{fig:korb} illustrates $\Delta K^{\rm orb}(j,n_e)$ as a function of $n_e$.
We find that the contribution from the diamagnetic current operator to the density-dependent part of the Knight shift, $\Delta K^{\rm orb}$,  is negligible.
However, the paramagnetic current operator leads to a  $\Delta K^{\rm orb}$ that is of the same order of magnitude as $\Delta K^{\rm cont}$ and $\Delta K^{\rm dip}$.
Accordingly, orbital currents play an important role in the density-dependence of the NMR shifts in \bise\, and \bite.
This result is unexpected, because the orbital shift is frequently believed to have a much weaker dependence on the carrier concentration than the contact and dipolar shifts.
Also, we find $\Delta K^{\rm orb}(j_0,n_e)>0$ for $n_e>0$ and $\Delta K^{\rm orb}(j_0, n_e)<0$ for $n_e<0$, which means that the orbital shift tends to become more paramagnetic as the number of itinerant electrons increases, even though the total shift $K^{\rm orb}(j_0,n_e)$ (not shown) is often diamagnetic.
In absence of spin-orbit interactions, the orbital shift is significantly weakened but remains anisotropic ($\Delta K_{xx}^{\rm orb}\neq \Delta K_{zz}^{\rm orb}$) because of the layered crystal structure.

%%%%%%%%%%%%%%%%%%%%%%%                                                                              
\begin{figure}[t]
\centering
\includegraphics[width=\columnwidth]{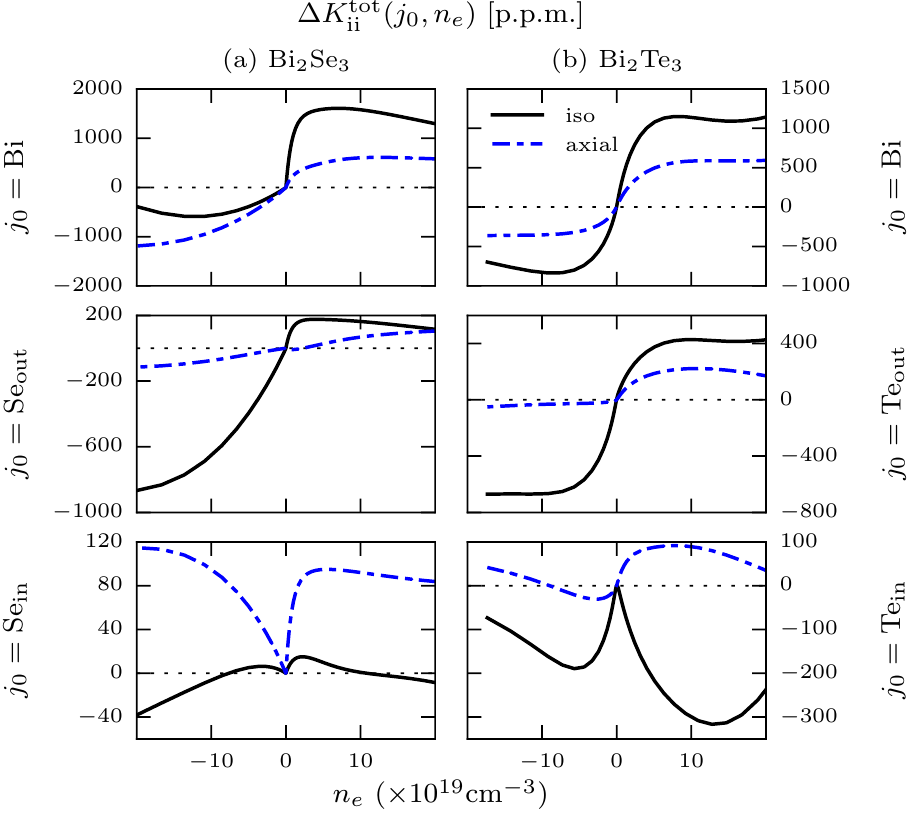}
\caption{Total (sum of Eqs.~(\ref{eq:dk}) and (\ref{eq:dk2})) room-temperature NMR shift for \bise\, and \bite, in parts per million,  as a function of carrier density $n_e$.
The ``iso'' curves show the trace of the total NMR shift, $(\Delta K^{\rm tot}_{zz}+2 \Delta K^{\rm tot}_{xx})/3$.
The ``axial'' curves show the anisotropic part of the shift, $(\Delta K^{\rm tot}_{zz}-\Delta K^{\rm tot}_{xx})/3$.}
\label{fig:total_iso}
\end{figure}
%%%%%%%%%%%%%%%%%%%%%%%%%%

\section{Discussion and Conclusions}\label{sec:discussion}

In summary, we have presented a $sp^3$ tight-binding theory of NMR shifts in two model topological insulators, \bise\, and \bite.
Specifically, we have focused on the density-dependent parts of the contact, dipolar, core polarization and orbital shifts, for which our theory is more reliable.
This part of the NMR shift can be accessed by doing measurements on samples with different carrier concentrations.
Along the way, we have introduced (new, to our knowledge) gauge-invariant expressions for the local spin and current densities (e.g. Eq.~\ref{eq:S2}), which can also be used to evaluate NMR shifts in materials other than \bise\, and \bite.
The tight-binding approach has enabled us to obtain the first theoretical estimates of the hyperfine couplings for different nuclei in  \bise\, and \bite, as well as to evaluate the $g$ factors.
These results could not have been attained with a minimal ${\bf k}\cdot{\bf p}$ model describing the electronic structure in the vicinity of the $\Gamma$ point (cf. the Introduction).
Granted, the tight-binding approach is not a full-blown first-principles approach.
Fully first-principles studies of carrier density-dependent NMR shifts in strongly spin-orbit coupled crystals with complex lattice structures have not been yet completed. 
Thus, the tight-binding method strikes a compromise allowing to obtain realistic estimates of the carrier density-dependent portion of the NMR shifts in topological materials with a modest computational effort. 
In addition, the tight-binding formalism is versatile enough that it will enable to calculate the topological surface state contributions to NMR in thin films and nanowires.\cite{nisson2014}
Considering that the density-independent chemical shift can be efficiently calculated from first-principles,\cite{wien2k,Njegic,zhang2016}
an interesting research direction would be to combine the tight-binding approach to calculate the density-dependent part of the NMR shift, with ab-initio methods to calculate the chemical shift contribution from the core electrons. 

The experimentally measured NMR shift is the sum of all individual parts (contact, dipolar, core, orbital).
Our calculated values for the density-dependent part of the total shift are shown in Fig.~\ref{fig:total_iso}.
In particular, this figure displays the isotropic and anisotropic components of the shift, i.e. $\Delta K_{\rm iso}= (\Delta K^{\rm tot}_{z z} + 2 \Delta K^{\rm tot}_{x x})/3$ and $\Delta K_{\rm axial}=(\Delta K^{\rm tot}_{zz}-\Delta K^{\rm tot}_{xx})/3$. 
Some thought is required when comparing our theoretical results to experiment.
At weak fields, the resonance frequency for a nucleus $j_0$ measured in a sample with carrier density $n_e$ can be written as
\begin{equation}
\omega(j_0, n_e) = \gamma(j_0) \left[1+K(j_0,n_e)\right] B,
\end{equation}
where $\gamma(j_0)$ is the nuclear magnetogyric ratio and $B$ is the external magnetic field and $K$ is the total shift (we drop the superscript ``tot'' for brevity).
Nowak\cite{nowak2015} has pointed out that, in $^{209}$Bi NMR, the experimentally reported values\cite{young2012, nisson2013} of the {\em absolute} shifts $K(j_0, n_e)$ 
may vary in magnitude and even in sign depending on the chosen standard for $\gamma(j_0)$ and the shift reference.
While this problem affects largely the chemical shift, it becomes much less severe for the {\em relative} shift $\Delta \omega(j_0, n_e) = \omega(j_0, n_e) - \omega(j_0, 0)$, whose sign will not depend on the choice of the standard (the magnitude of the shift will depend slighly on the standard).
By construction, our theoretical value for $\Delta K(j_0,n_e)$ is independent of $\gamma(j_0)$, and is equivalent to the experimentally measurable quantity $\Delta\omega(j_0,n_e)/(\gamma(j_0) B)$.
Admittedly, in bulk crystals it is not easy to control the carrier density, and thus it may be a challenge to measure $\omega(j_0,0)$ directly.
However, one is often able to measure the shift at two or more different carrier concentrations (e.g. $n_1$ and $n_2$, not necessarily vanishing).
Then,  $[\omega(j_0,n_1)-\omega(j_0,n_2)]/(\gamma(j_0) B)$ is identical to $\Delta K(j_0, n_1)-\Delta K(j_0, n_2)$ and can thus be compared to our theory.  
%%%%%%%%%%%%%%%%%%%%%%%%%%                                                     
\begin{figure}[tb]
\centering
\includegraphics[width=1\columnwidth]{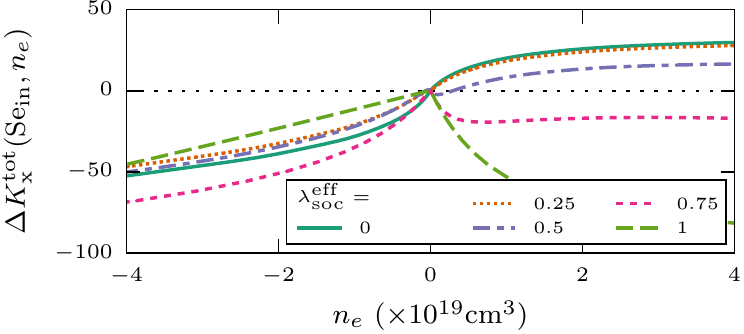}
\caption{The evolution of $\Delta K_{\rm xx}($\Sei$, n_e)$ across a band inversion. 
The atomic spin-orbit parameters are scaled by hand in such a way that the bandgap closes at $\lambda_{\rm soc}^{\rm eff} \sim 0.5$,  $\lambda_{\rm soc}^{\rm eff}>0.5$ corresponds to a topological insulator and $\lambda_{\rm soc}^{\rm eff} < 0.5$ corresponds to a trivial insulator. The value $\lambda_{\rm soc}^{\rm eff} =1$ corresponds to the actual spin-orbit parameters of \bise, used in the previous figures. }
\label{fig:inversion}
\end{figure}
 %%%%%%%%%%%%%%%%%%%%%%%%%%                                                     

Overall, the picture drawn from Fig.~\ref{fig:total_iso} is rather complex: Bi shifts are dominated by the contact interaction, \Seo\, and \Teo\, shifts are dominated by orbital currents, the \Sei\, shift is dominated by dipolar interactions and the \Tei\, shift contains a close competition between contact, dipolar, orbital and core polarization contributions.
The dominant role played by the contact interaction in $^{209}$ Bi corroborates the interpretation of the experiments by Young {\em et al.}\cite{young2012} and by
Mukhopadhyay {\em et al.}\cite{mukhopadhyay2015}.
The reason why the contact interaction is relatively more important in Bi has to do with the fact that it is heavier and that the low-energy bands have a higher $s-$orbital projection on Bi sites.
The calculated $\Delta K_{\rm iso}({\rm Bi}, n_e)\simeq 0.2\%$ for $n_e\simeq 5\times 10^{19}{\rm cm}^{-3}$ is also in quite good agreement with experiment.\cite{nisson2013} (see also the Supplemental Material of Ref.~[\onlinecite{mukhopadhyay2015}]).

With the exception of \Sei\, and \Tei, we find that $\Delta K_{\rm iso}(j_0,n_e)$ is positive for $n-$doped samples and negative for $p-$doped samples.
This is consistent with the experiment from Ref.~[\onlinecite{young2012}], as well as with the statement\cite{nowak2014} that ``for classical semiconductors, the sign of the Knight shift measured relative to the resonance of a carrier-free sample is negative for hole-doped materials and positive for electron-doped materials''.
At low carrier densities and at room temperature, $\Delta K_{\rm iso}$ and $\Delta K_{\rm axial}$ vary linearly with $|n_e|$,
which is expected in a non-degenerate Fermi gas.\cite{selbach1979}
At higher $|n_e|$, $\Delta K_{\rm iso}$ and $\Delta K_{\rm aniso}$ become non monotonic, a result that is reminiscent to the one measured\cite{hewes1973} in Pb$_x$Sn$_{1-x}$Te.
In electron-doped \bise, at carrier densities not exceeding $\simeq 5\times 10^{19} {\rm cm}^{-3}$,  $|\Delta K_{\rm iso}|/|\Delta K_{\rm axial}|$ grows with $n_e$, in apparent agreement with the experimental observation in the Supplemental Material of Ref.~[\onlinecite{mukhopadhyay2015}]. 
Another result from our theory, worth stressing because it has been generally overlooked in the experiments, is that dipolar interactions can make a large contribution to the {\em isotropic} NMR shift, while the contact interaction can make a large contribution to the {\em anisotropic} shift. 
This is a consequence of strong spin-orbit interactions in \bise\, and \bite. 
 
Finally, motivated by recent suggestions\cite{nowak2014, shi2014, zhang2016} that the NMR shift could contain systematic and possibly universal differences between topological and trivial insulators,
we compute $\Delta K(j_0,n_e)$ for different values of the spin-orbit coupling.
As an example, Fig.~\ref{fig:inversion} shows $\Delta K_{\rm xx}($\Sei$, n_e)$ for various scalings of the spin-orbit coupling by the dimensionless factor $\lambda_{\rm soc}^{\rm eff}$.
Analogous curves for other nuclei are not shown.
We observe a sign change of $\Delta K_{\rm xx}($\Sei$,n_e)$ for electron-doped systems close to the spin-orbit strength where the gap closes $\lambda_{\rm soc}^{\rm eff} \sim 0.5$, i.e. close to the topological phase transition.
A similar sign change takes place for $\Delta K_{\rm xx}($\Tei$,n_e)$ in the case of hole-doped \bite.
However, these trends are neither universal nor correlated with a topological invariant; instead, they simply 
evidence the major impact that spin-orbit interaction has in the density-dependence of the NMR shift.
Likewise, we believe that the difference observed in Refs.~[\onlinecite{nowak2014,shi2014}] between the shifts of YPdBi and YPtBi is a reflection of the particular orbital character of the low energy bands in these specific materials, as opposed to a universal feature associated to topological invariants. 
At any rate, more theoretical research is needed before generally ruling out genuine topological signatures from bulk NMR. 
For instance, based on Ref.~[\onlinecite{saha2015}], it might be that the carrier-density-dependence of bulk NMR {\em linewidths} will display fingerprints of band inversions.  

In conclusion, we have presented a theory of NMR shifts in model topological insulators.
This work has been motivated by an array of recent NMR experiments in \bise\, and \bite, and hopes to help the interpretation of upcoming ones in related materials. 
Possible avenues for future research include applying our formalism to thin films in order to calculate the NMR shifts as a function of the film thickness (which would incorporate both bulk and surface state contributions), searching for possible band inversion effects in the  $T_1$ relaxation time, reassessing the importance of the Bloembergen-Rowland interactions in the large magnetic-field-independent NMR linewidths,\cite{georgieva2016} performing a theoretical modelling of the NMR anomalies\cite{matano2016} in the superconducting phase of Cu-doped \bise,
and applying the formalism presented here to other topological materials.

\begin{acknowledgments}
IG is indebted to N. Georgieva and J. Haase for helpful conversations and for the collaboration in Ref.~[\onlinecite{georgieva2016}], which provided the impetus for the present work.
JRR benefited from discussions with K. Saha in the early stages of the work, and was supported by the government of Canada through a Mitacs Globalink undergraduate research fellowship. 
SB and IG thank Qu\'ebec's RQMP and Canada's NSERC for funding, Calcul Qu\'ebec and Compute Canada for computer resources, and
J. Quilliam for discussions.
\end{acknowledgments}

\appendix

\begin{widetext}
\section{Derivation of Eq.~(\ref{eq:S2})}\label{app:derivationPT}
In this Appendix, we use perturbation theory to derive Eq.~(\ref{eq:S2}) of the main text.
This equation allows to calculate the expectation values of local operators 
$\langle O({\bf r})\rangle$
encountered in the calculation of NMR shifts, with ${\bf O}({\bf r}) = {\bf S}({\bf r})$ for the Knight shift and ${\bf O}({\bf r})={\bf J}({\bf r})$ for the orbital shift.
These expectation values can be expressed in the eigenbasis of the full Hamiltonian, $\{|\psi_i\rangle\}$, via
\begin{equation}
\langle{\bf O}({\bf r})\rangle = \sum_i \langle\psi_i | {\bf O}({\bf r}) | \psi_i \rangle f_i,
\end{equation}
where $f_i$ is the Fermi occupation factor for the state $|\psi_i\rangle$ with energy $E_i$.
In linear response,  we expand ${\bf O}({\bf r})$, $|\psi_i\rangle$ and $E_i$ to first order\cite{pert} in the perturbation $\delta{\cal H}=\mu_B ({\boldsymbol\sigma}+m {\bf r}\times{\bf v}/\hbar)\cdot{\bf B}$,
\begin{equation}
\label{eq:tr}
\langle{\bf O}({\bf r})\rangle \simeq \sum_i\langle\psi_i^{(0)} | {\bf O}^{(0)}({\bf r}) | \psi_i^{(0)}\rangle f_i^{(0)} + \sum_i \langle\psi_i^{(0)} | \delta{\bf O}({\bf r}) | \psi_i^{(0)}\rangle f_i^{(0)} + \sum_{i j} \langle\psi_i^{(0)}|{\bf O}^{(0)}|\psi_j^{(0)}\rangle \langle\psi_j^{(0)}|\delta{\cal H}|\psi_i^{(0)}\rangle \frac{f_i^{(0)}-f_j^{(0)}}{E_i^{(0)}-E_j^{(0)}}.
\end{equation}
Here, the superscript ``(0)'' refers to the unperturbed case with $\delta{\cal H}=0$, and $\delta {\bf O}$ denotes the term linear in $\delta{\cal H}$ when expanding ${\bf O}$.
In our case, $\delta {\bf S}({\bf r})=\delta{\bf J}_P({\bf r})=0$ but $\delta {\bf J}_D({\bf r}) \neq 0$ .
Also, $\sum_i\langle\psi_i^{(0)} | {\bf O}^{(0)}({\bf r}) | \psi_i^{(0)}\rangle f_i^{(0)} = 0$ for our cases of interest.
Incidentally, $i=j$ terms in $\sum_{i j}$ (which originate from the changes in the occupation numbers due to the perturbation) must be understood via $(f_i^{(0)}-f_j^{(0)})/(E_i^{(0)}-E_j^{(0)})\to \partial f_i^{(0)}/\partial E_i^{(0)}$.
In addition, in this appendix we shall ignore $\delta{\bf O}^{(0)}({\bf r})$, which is nonzero only for the the diamagnetic current density (cf. App.~\ref{subsec:dia} for a discussion of this contribution).
With this proviso, in the band eigenstate basis where $|\psi^{(0)}_i\rangle = |\psi_{{\bf k} n}\rangle$, with $|\psi_{{\bf k} n}\rangle$ the eigenstates of Eq.~(\ref{eq:h0}), Eq.~(\ref{eq:tr}) reads
\begin{equation}
\label{eq:tr2}
\langle{\bf O}({\bf r})\rangle=\sum_{{\bf k}{\bf k}'}\sum_{n n'}\langle\psi_{{\bf k} n}|{\bf O}({\bf r})|\psi_{{\bf k}'n'}\rangle\langle\psi_{{\bf k}'n'} | \delta{\cal H} |\psi_{{\bf k} n}\rangle f_{nn'}({\bf k},{\bf k}'), \text{    (for ${\bf O}({\bf r})={\bf S}({\bf r}), {\bf J}_P({\bf r})$)}
\end{equation}
where $f_{nn'}({\bf k},{\bf k}')$ was defined in Eq.~(\ref{eq:fnn}).
For brevity of notation, we have omitted the superscript ``(0)'' for ${\bf O}$ in the right hand side of Eq.~(\ref{eq:tr2}).
We will keep this omission for the remainder of the appendix.  

Herein, we are interested in the matrix elements of $\delta{\cal H}$,
\begin{equation}
\label{eq:hz2}
\langle\psi_{{\bf k}'n'}|\delta{\cal H}|\psi_{{\bf k} n}\rangle = \mu_B \delta_{{\bf k} {\bf k}'} \langle\psi_{{\bf k}'n'}|{\boldsymbol\sigma}\cdot{\bf B} |\psi_{{\bf k} n}\rangle+\frac{m\mu_B}{\hbar}\langle\psi_{{\bf k}'n'}|({\bf r}\times{\bf v})\cdot{\bf B} |\psi_{{\bf k} n}\rangle.
\end{equation}
The first term on the right hand side of Eq.~(\ref{eq:hz2}) is diagonal in crystal momentum and straightforward to compute.
The second term is more delicate.
Starting from
\begin{equation}
\label{eq:a5}
\langle \psi_{{\bf k}' n'}|{\bf r}\times{\bf v}|\psi_{{\bf k} n}\rangle 
= \sum_{{\bf k}'' n''} \langle\psi_{{\bf k}' n'}|{\bf r}|\psi_{{\bf k}'' n''}\rangle\times\langle\psi_{{\bf k}'' n''}| {\bf v} | \psi_{{\bf k} n}\rangle\
=\sum_{n''} \langle\psi_{{\bf k}' n'}|{\bf r}|\psi_{{\bf k} n''}\rangle\times\langle\psi_{{\bf k} n''}| {\bf v} | \psi_{{\bf k} n}\rangle,
\end{equation}
and using (cf. Ref.[\onlinecite{blount1962}])
\begin{equation}
\langle\psi_{{\bf k}'n'}|{\bf r}|\psi_{{\bf k} n}\rangle = (1/N) \int \drm^3 r (-i) (\grad_{\bf k} e^{i {\bf k}\cdot{\bf r}}) e^{-i{\bf k}'\cdot{\bf r}} u^*_{{\bf k}'n'}({\bf r}) u_{{\bf k} n}({\bf r})=-i\delta_{n n'}\grad_{\bf k}(\delta_{{\bf k}{\bf k}'}) + i \delta_{{\bf k}{\bf k}'} \langle u_{{\bf k} n'}|\grad_{\bf k} u_{{\bf k} n}\rangle_{\rm cell},
\end{equation}
we obtain
\begin{align}
\label{eq:orb2}
\langle \psi_{{\bf k}' n'}|{\bf r}\times{\bf v}|\psi_{{\bf k} n}\rangle=
-\frac{i}{\hbar} (\grad_{\bf k} \delta_{{\bf k}{\bf k}'})\times \langle u_{{\bf k} n'}|(\grad_{\bf k}{\cal H}_{\bf k})|u_{{\bf k} n}\rangle_{\rm cell}
 -\frac{i}{\hbar}\delta_{{\bf k}{\bf k}'} \langle \grad_{\bf k} u_{{\bf k} n'} |\times (\grad_{\bf k} {\cal H}_{\bf k})|u_{{\bf k} n}\rangle_{\rm cell},
\end{align}
where ${\cal H}_{\bf k} = e^{-i {\bf k}\cdot{\bf r}} {\cal H} e^{i {\bf k}\cdot{\bf r}}$, $|u_{{\bf k} n}\rangle = \sqrt{N} e^{-i {\bf k}\cdot{\bf r}} |\psi_{{\bf k} n}\rangle$, and $ \grad_{\bf k}{\cal H}_{\bf k}=\hbar\, e^{-i{\bf k}\cdot{\bf r}} {\bf v} e^{i {\bf k}\cdot{\bf r}} $.
In the derivation of Eq.~(\ref{eq:orb2}), we have used $\hbar \langle\psi_{{\bf k} n}|{\bf v}|\psi_{{\bf k}n'}\rangle=\langle u_{{\bf k} n}|(\grad_{\bf k}{\cal H}_{\bf k})|u_{{\bf k} n'}\rangle_{\rm cell}$, $\langle u_{{\bf k} n'}|\grad_{\bf k} u_{{\bf k} n}\rangle_{\rm cell}=-\langle \grad_{\bf k} u_{{\bf k} n'}|u_{{\bf k} n}\rangle_{\rm cell}$, and $\sum_{n''} |u_{{\bf k} n''}\rangle\langle u_{{\bf k} n''}|={\bf 1}$.
Also, $\langle\,| ...|\,\rangle_{\rm cell}$ means that the spatial integral is carried out within a unit cell.
Although not immediately obvious from Eq.~(\ref{eq:orb2}), it can be shown explicitly that $\langle\psi_{{\bf k}' n'} |{\bf r}\times{\bf v}|\psi_{{\bf k} n}\rangle^*=\langle\psi_{{\bf k} n} |{\bf r}\times{\bf v}|\psi_{{\bf k}' n'}\rangle$. 
Taking advantage of this, it is convenient to rewrite Eq.~(\ref{eq:orb2}) in a more symmetric form,
\begin{align}
\label{eq:orb3}
\begin{split}
&\langle \psi_{{\bf k}' n'}|{\bf r}\times{\bf v}|\psi_{{\bf k} n}\rangle=\frac{1}{2}\left(\langle \psi_{{\bf k}' n'}|{\bf r}\times{\bf v}|\psi_{{\bf k} n}\rangle +\langle \psi_{{\bf k} n}|{\bf r}\times{\bf v}|\psi_{{\bf k}' n'}\rangle^*\right)=\\
&-\frac{i}{2\hbar} (\grad_{\bf k} \delta_{{\bf k}{\bf k}'}) \times \langle u_{{\bf k} n'}|(\grad_{\bf k} {\cal H}_{\bf k})|u_{{\bf k} n}\rangle_{\rm cell}
+\frac{i}{2\hbar} (\grad_{{\bf k}'}  \delta_{{\bf k}{\bf k}'}) \times \langle u_{{\bf k}' n'}|(\grad_{{\bf k}'} {\cal H}_{{\bf k}'})|u_{{\bf k}' n}\rangle_{\rm cell}\\
&-\frac{i}{2\hbar}\delta_{{\bf k} {\bf k}'}\langle \grad_{\bf k} u_{{\bf k} n'} |\times (\grad_{\bf k} {\cal H}_{\bf k})|u_{{\bf k} n}\rangle_{\rm cell}
-\frac{i}{2\hbar}\delta_{{\bf k} {\bf k}'} \langle u_{{\bf k} n'} |(\grad_{\bf k} {\cal H}_{\bf k})\times |\grad_{\bf k} u_{{\bf k} n}\rangle_{\rm cell}.
\end{split}
\end{align}
Combining Eqs.~(\ref{eq:tr2}), (\ref{eq:hz2}) and (\ref{eq:orb3}), we arrive at
\begin{align}
\label{eq:oo}
\begin{split}
\langle O_i({\bf r})\rangle &=\mu_B \sum_{{\bf k}}\sum_{ n n'} \langle\psi_{{\bf k} n} | O_i({\bf r}) | \psi_{{\bf k} n'}\rangle \langle \psi_{{\bf k} n'} |{\boldsymbol\sigma}\cdot{\bf B} |\psi_{{\bf k} n}\rangle f_{n n'}({\bf k},{\bf k})\\
&+\frac{m \mu_B}{\hbar^2} {\rm Im} \sum_{\bf k}\sum_{n n'} \langle\psi_{{\bf k} n} | O_i({\bf r}) | \psi_{{\bf k} n'}\rangle f_{n n'}({\bf k},{\bf k}) \left(\langle\grad_{\bf k} u_{{\bf k} n'} | \times (\grad_{\bf k} {\cal H}_{\bf k}) |u_{{\bf k} n}\rangle_{\rm cell}\right)\cdot{\bf B}\\\
&-\frac{m\mu_B}{\hbar} {\rm Im} \sum_{{\bf k}{\bf k}'} \sum_{n n'} \delta_{{\bf k}{\bf k}'} \left[\grad_{\bf k}\times\left(\langle\psi_{{\bf k} n} | O_i({\bf r}) | \psi_{{\bf k}' n'}\rangle  f_{n n'}({\bf k},{\bf k}') 
\langle \psi_{{\bf k} n'}|{\bf v} | \psi_{{\bf k} n}\rangle\right)\right]\cdot{\bf B},
\end{split}
\end{align}
where $i=x,y,z$ and ${\rm Im}$ stands for the imaginary part.
Both ${\bf k}$ and ${\bf k}'$ are within the first Brillouin zone.
In the numerical calculations, we use a Monkhorst-Pack mesh\cite{monkhorst1976} with $100\times 100\times 100$ $k$ points.
In the derivation of Eq.~(\ref{eq:oo}) we have used 
\begin{equation}
\label{eq:deriv}
\sum_{{\bf k}{\bf k}'}  \grad_{\bf k} (\delta_{{\bf k} {\bf k}'}) F({\bf k}, {\bf k}') = -\sum_{{\bf k}{\bf k'}} \delta_{{\bf k}{\bf k}'} \grad_{\bf k} F({\bf k},{\bf k}'), 
\end{equation}
where $F({\bf k}, {\bf k}')$ is a differentiable function of ${\bf k}$ and ${\bf k}'$, periodic in the reciprocal space.
In the last term on the right hand side of Eq.~(\ref{eq:oo}), one must take the derivative with respect to ${\bf k}$ {\em before} applying $\delta_{{\bf k} {\bf k}'}$.
It can be verified explicitly that $\langle {\bf O} ({\bf r})\rangle$ is invariant under gauge transformations.

Equation~(\ref{eq:oo}) is one of the main results of this appendix.
It incorporates interband transitions ($E_{{\bf k} n}\neq E_{{\bf k} n'}$) and remains valid when the Fermi energy is far from a band extremum. 
We use it to compute the Knight shift and the paramagnetic orbital shift (the diamagnetic orbital shift may be evaluated directly from the $\delta{\bf O}({\bf r})$ term in Eq.~(\ref{eq:tr2}), which is relatively easy to treat, cf. Appendix~\ref{app:orbital}).
In the numerical evaluation of Eq.~(\ref{eq:oo}), $\grad_{\bf k}\times$ in the last term and $\langle\grad_{\bf k} u_{{\bf k} n'}|$ in the second term are evaluated by discretizing, e.g.
\begin{equation}
\sum_{{\bf k}{\bf k'}} \delta_{{\bf k}{\bf k}'} \grad_{\bf k}\times {\bf F}({\bf k},{\bf k}')=\lim_{q\to 0}\frac{1}{2 q}\sum_{\bf k}\sum_{\alpha}\hat{\bf e}_\alpha 
\times\left[{\bf F}({\bf k}+q \hat{\bf e}_\alpha, {\bf k}) - (q\to -q)\right],
\end{equation}
where $\hat{\bf e}_\alpha$ is a unit vector ($\alpha=x,y,z$) and ${\bf F}({\bf k},{\bf k}')$ is a vector differentiable with respect to its arguments. 
Thereafter we adopt a smooth gauge, such that $|\psi_{{\bf k}+q \hat{\bf e}_\alpha n}\rangle \to |\psi_{{\bf k} n}\rangle$ as $q\to 0$, which is possible to do in \bise\, and \bite\, because they have zero Chern number.
In this regards, it is useful to split the double degeneracy of each band by a very small Zeeman splitting.
In addition, we employ the relations $\sum_n |u_{{\bf k} n}\rangle\langle u_{{\bf k} n}|={\bf 1}$ and
\begin{equation}
\langle u_{{\bf k}' n'} |(\grad_{\bf k} {\cal H}_{\bf k})|u_{{\bf k} n}\rangle_{\rm cell}= \hbar \langle\psi_{{\bf k}' n'}|e^{i({\bf k}'-{\bf k})\cdot{\bf r}} {\bf v}|\psi_{{\bf k} n}\rangle
\simeq\sum_{j\mu\sigma}\sum_{j'\mu'\sigma'} C^*_{{\bf k}' n'; {\bf k}' j' \mu' \sigma'} C_{{\bf k} n; {\bf k} j \mu \sigma}\grad_{\bf k} {\cal H}_{j'\mu'\sigma', j \mu\sigma}({\bf k}),
\end{equation}
where we have neglected the intra atomic dipole matrix element in the last equality.

Finally, using the closure relation $\sum_n |u_{{\bf k} n}\rangle\langle u_{{\bf k} n}|={\bf 1}$, as well as the relation 
\begin{equation}
\langle \grad_{\bf k} u_{{\bf k} n}|u_{{\bf k} n'}\rangle_{\rm cell} = \langle u_{{\bf k} n}| (\grad_{\bf k} {\cal H}_{\bf k})|u_{{\bf k} n'}\rangle_{\rm cell}/(E_{{\bf k} n}-E_{{\bf k} n'})
\qquad
(\text{when } E_{{\bf k} n}\neq E_{{\bf k} n'}), 
\end{equation}
and upon replacing ${\bf O}$ by ${\bf S}$, Eq.~(\ref{eq:oo}) becomes Eq.~(\ref{eq:S2}) of the main text.

\section{Details concerning the calculation of the dipolar shift}\label{app:dipolar}

In this appendix we present some details of the calculation of the dipolar Knight shift, which are important in order to obtain Fig.~\ref{fig:kdip}.
The objective is to evaluate
\begin{equation}
\label{eq:dibat}
\delta{\bf B}^{\rm dip}({\bf r}_0)=\frac{\mu_0}{4\pi} g_s \mu_B \int \drm^3 r \frac{1}{|{\bf r}-{\bf r}_0|^3} \left[\langle{\bf S}({\bf r})\rangle - 3 \frac{{\bf r}-{\bf r}_0}{|{\bf r}-{\bf r}_0|} \langle{\bf S}({\bf r})\rangle\cdot\frac{{\bf r}-{\bf r}_0}{|{\bf r}-{\bf r}_0|}\right]
\end{equation}
numerically.
The expectation value $\langle {\bf S} ({\bf r}) \rangle$ can be calculated from Eq.~\eqref{eq:oo}, which requires the matrix elements of the spin density operator in the band eigenstate basis. Those are given by
\begin{equation}
\langle\psi_{{\bf k} n}|S_i({\bf r})|\psi_{{\bf k}'n'}\rangle \simeq \frac{1}{2 N} \sum_{{\bf R} j \mu\sigma}\sum_{\mu'\sigma'} e^{i({\bf k}'-{\bf k})\cdot({\bf R}+{\bf t}_j)} \langle\sigma|{\boldsymbol\sigma}|\sigma'\rangle \langle\psi_{{\bf k} n}|\psi_{{\bf k} j \mu \sigma}\rangle\langle\psi_{{\bf k}'j\mu'\sigma'}|\psi_{{\bf k}'n'}\rangle \langle{\bf r}|{\bf R} j \mu\rangle\langle{\bf r}|{\bf R} j \mu'\rangle,
\end{equation}
where we have exploited the fact that the atomic wave functions are real and exponentially localized.
Because the integrand of Eq.~(\ref{eq:dibat}) is peaked at ${\bf r}\simeq {\bf r}_0$, the main contribution to $\delta{\bf B}^{\rm dip}({\bf r}_0)$ comes from the spin density in the vicinity of ${\bf r}_0$. 
Longer range contributions are neglected.
This leads us to further approximate 
\begin{equation}
\langle\psi_{{\bf k} n}|S_i({\bf r})|\psi_{{\bf k}'n'}\rangle \simeq \frac{1}{2 N}  e^{i({\bf k}'-{\bf k})\cdot {\bf r}_0} \sum_{\mu\sigma}\sum_{\mu'\sigma'} \langle\sigma|{\boldsymbol\sigma}|\sigma'\rangle \langle\psi_{{\bf k} n}|\psi_{{\bf k} j_0 \mu \sigma}\rangle\langle\psi_{{\bf k}'j_0\mu'\sigma'}|\psi_{{\bf k}'n'}\rangle \langle{\bf r}|{\bf R}_0 j_0 \mu\rangle\langle{\bf r}|{\bf R}_0 j_0 \mu'\rangle,
\end{equation}
where we have taken ${\bf R} + {\bf t}_j \simeq {\bf r}_0$.
Consequently, the spatial integral that we need to compute is
\begin{align}
\label{eq:dihir}
\begin{split}
&\int \drm^3 r \frac{1}{|{\bf r}-{\bf r}_0|^3}\left[\langle\psi_{{\bf k} n}|{\bf S}({\bf r})|\psi_{{\bf k}'n'}\rangle - 3 \frac{{\bf r}-{\bf r}_0}{|{\bf r}-{\bf r}_0|} \langle\psi_{{\bf k} n}|{\bf S}({\bf r})|\psi_{{\bf k}'n'}\rangle\cdot\frac{{\bf r}-{\bf r}_0}{|{\bf r}-{\bf r}_0|}\right]\\
&\simeq\frac{1}{2 N} e^{i({\bf k}'-{\bf k})\cdot{\bf r}_0} \sum_{\mu\mu'}\sum_{\sigma\sigma'} \langle\psi_{{\bf k} n}|\psi_{{\bf k} j_0 \mu\sigma}\rangle\langle\psi_{{\bf k}' j_0 \mu'\sigma'}|\psi_{{\bf k}' n}\rangle \\
&~~~~~~~~~~~~\int \drm^3 r \frac{1}{|{\bf r}-{\bf r}_0|^3}\left[\langle\sigma|{\boldsymbol\sigma}|\sigma'\rangle - 3 \frac{{\bf r}-{\bf r}_0}{|{\bf r}-{\bf r}_0|}\langle\sigma|{\boldsymbol\sigma}|\sigma'\rangle\cdot\frac{{\bf r}-{\bf r}_0}{|{\bf r}-{\bf r}_0|}\right]\langle{\bf r}|{\bf R}_0 j_0 \mu\rangle\langle{\bf r}|{\bf R}_0 j_0 \mu'\rangle.
\end{split}
\end{align}
In particular, let us concentrate on
\begin{align}
\label{eq:dihir2}
\begin{split}
&\int \drm^3 r \frac{1}{|{\bf r}-{\bf r}_0|^3}\left[\langle\sigma|{\boldsymbol\sigma}|\sigma'\rangle - 3 \frac{{\bf r}-{\bf r}_0}{|{\bf r}-{\bf r}_0|} \langle\sigma|{\boldsymbol\sigma}|\sigma'\rangle\cdot\frac{{\bf r}-{\bf r}_0}{|{\bf r}-{\bf r}_0|}\right]\langle{\bf r}|{\bf R}_0 j_0 \mu\rangle\langle{\bf r}|{\bf R}_0 j_0 \mu'\rangle\\
&=\int \drm^3 r \frac{1}{r^3}\left[\langle\sigma|{\boldsymbol\sigma}|\sigma'\rangle - 3 \hat{\bf r}\langle\sigma|{\boldsymbol\sigma}|\sigma'\rangle\cdot\hat{\bf r}\right]
\langle{\bf r}|{\bf 0} j_0 \mu\rangle\langle{\bf r}|{\bf 0} j_0 \mu'\rangle,
\end{split}
\end{align}
where the transition from the first to the second line follows from the fact that $\langle{\bf r}|{\bf R}_0 j_0 \mu\rangle$ is a function of ${\bf r}-{\bf R}_0-{\bf t}_{j_0}={\bf r}-{\bf r}_0$.
By definition, $\langle{\bf r}|{\bf 0} j_0 \mu\rangle$ is the wave function corresponding to the orbital $\mu$ of atom $j_0$, the atomic center being at the origin of the coordinate system.

The $z$ component of Eq.~(\ref{eq:dihir2}) can be written as
\begin{align}
\label{eq:dilau}
\int\! \drm^3 r \frac{1}{r^3}\left[\langle\sigma|\sigma^z|\sigma'\rangle - 3 \frac{z}{r}\langle\sigma|{\boldsymbol\sigma}|\sigma'\rangle\cdot\hat{\bf r}\right]\langle{\bf r}|{\bf 0} j_0 \mu\rangle\langle{\bf r}|{\bf 0} j_0 \mu'\rangle
&=\delta_{\mu\mu'}\langle\sigma|\sigma^z|\sigma'\rangle \int \drm^3 r\frac{1}{r^3}\left(1-\frac{3 z^2}{r^2}\right)\langle{\bf r}|{\bf 0} j_0 \mu\rangle^2\\
&-3 (\delta_{\mu p_z}\delta_{\mu' p_x}+\delta_{\mu' p_z}\delta_{\mu p_x}) \langle\sigma|\sigma^x|\sigma'\rangle\int \drm^3 r \frac{z x}{r^5}\langle{\bf r}|{\bf 0} j_0 p_x\rangle\langle{\bf r}|{\bf 0} j_0 p_z\rangle\nonumber\\
&-3 (\delta_{\mu p_z}\delta_{\mu' p_y}+\delta_{\mu' p_z}\delta_{\mu p_y}) \langle\sigma|\sigma^y|\sigma'\rangle\int \drm^3 r \frac{z y}{r^5}\langle{\bf r}|{\bf 0} j_0 p_y\rangle\langle{\bf r}|{\bf 0} j_0 p_z\rangle.\nonumber
\end{align}

Combining Eqs.~(\ref{eq:dibat}), (\ref{eq:oo}), (\ref{eq:dihir}) and (\ref{eq:dilau}), we arrive at
\begin{align}
\label{eq:dibost}
\begin{split}
&\delta B^{\rm dip}_z({\bf r}_0)
\simeq 
-\frac{1}{V}\sum_{{\bf k} n n'}
\langle\psi_{{\bf k} n}|
 M^z_{\kv, \kv} (j_0)|\psi_{{\bf k} n'}\rangle f_{n n'}({\bf k},{\bf k})\langle\psi_{{\bf k} n'}|{\boldsymbol\sigma}\cdot{\bf B}|\psi_{{\bf k} n}\rangle\\
&-\frac{m}{\hbar^2} {\rm Im}\frac{1}{V}\sum_{{\bf k} n n'}\langle\psi_{{\bf k} n}| M^z_{\kv,\kv}(j_0)|\psi_{{\bf k} n'}\rangle f_{n n'}({\bf k},{\bf k})
(\langle\grad_{\bf k} u_{{\bf k} n'}|\times(\grad_{\bf k}{\cal H}_{\bf k})|u_{{\bf k} n}\rangle_{\rm cell})\cdot{\bf B}\\
&+\frac{m}{\hbar^2} {\rm Im}\frac{1}{V}\sum_{{\bf k}{\bf k}' n n'}\delta_{{\bf k}{\bf k}'}
\left[\grad_{\bf k}\times\left(\langle\psi_{{\bf k} n}| M^z_{\kv,\kv'}(j_0)|\psi_{{\bf k}' n'}\rangle f_{n n'}({\bf k},{\bf k}')\langle\psi_{{\bf k} n'}|\hbar{\bf v}|\psi_{{\bf k} n}\rangle\right)\right]\cdot{\bf B},
\end{split}
\end{align}
where we have defined
\begin{align}
\label{eq:mz-dip}
M^z_{\kv,\kv'}(j_0)&=
\sum_{\mu \mu'} \sum_{\sigma \sigma'} \parS{
\beta(j_0)(\delta_{\mu p_x}\delta_{\mu' p_x}+\delta_{\mu p_y}\delta_{\mu' p_y}-2\delta_{\mu p_z}\delta_{\mu' p_z})\langle\sigma|\sigma^z|\sigma'\rangle
-3\beta' (j_0)(\delta_{\mu p_z}\delta_{\mu' p_x}+\delta_{\mu p_x}\delta_{\mu' p_z})\langle\sigma|\sigma^x|\sigma'\rangle
\right.
\nonumber \\ &\left. \qquad\qquad
-3\beta' (j_0)(\delta_{\mu p_z}\delta_{\mu' p_y}+\delta_{\mu p_y}\delta_{\mu' p_z})\langle\sigma|\sigma^y|\sigma'\rangle}
|\psi_{\kv j_0 \mu \sigma} \rangle \langle \psi_{\kv' j_0 \mu' \sigma'} |.
\end{align}
Here, $\delta_{ij}$ is Kronecker's delta, while $\beta(j_0)$ and $\beta'(j_0)$ are numerical factors computed using the wave functions calculated using atomic DFT\cite{psi4},
\begin{equation}
\label{eq:betanum}
\beta(j_0)=V_{\rm cell}\frac{\mu_0\mu_B^2}{4\pi} \int \drm^3 r\frac{1}{r^3}\left(1-\frac{3 z^2}{r^2}\right)\langle{\bf r}|{\bf 0} j_0 p_x\rangle^2
=\left\{\begin{array}{cc}  
0.382\, {\rm eV} \text{\AA}^3 & \text{ for $j_0$=Bi in \bise}\\
0.458\,{\rm eV} \text{\AA}^3 & \text{ for $j_0$=Bi in \bite}\\
0.222\,{\rm eV} \text{\AA}^3 & \text{ for $j_0$=Se in \bise}\\
0.333\,{\rm eV} \text{\AA}^3 & \text{ for $j_0$=Te in \bite}
\end{array}
\right.
\end{equation}
\begin{equation}
\beta'(j_0)=V_{\rm cell}\frac{\mu_0\mu_B^2}{4\pi}\int \drm^3 r \frac{z x}{r^5}\langle{\bf r}|{\bf 0} j_0 p_x\rangle\langle{\bf r}|{\bf 0} j_0 p_z\rangle
=\frac{1}{2}\beta(j_0).
\end{equation}
Note that $\beta({\rm Bi})$ takes slightly different values in \bise\, and \bite, because their unit cell volumes are not identical.
In the derivation of Eq.~(\ref{eq:dibost}), we have used $\mu_B = e \hbar/(2 m)$ and  $N=V/V_{\rm cell}$, where $V_{\rm cell}$ is the unit cell volume.
In addition, we have set $\exp[i({\bf k}'-{\bf k})\cdot{\bf r}_0] \to 1$.

Let us justify setting $\exp[i({\bf k}'-{\bf k})\cdot{\bf r}_0] \to 1$.
In general, we have an expression of the type
\begin{equation}
\sum_{{\bf k},{\bf k}'} \grad_{\bf k} \times\left[e^{i({\bf k}'-{\bf k})\cdot{\bf r}_0} {\bf F}({\bf k}, {\bf k}')\right]
=\sum_{\bf k}\lim_{q\to 0}\frac{1}{2 q}\sum_\alpha \hat{\bf e}_\alpha \times\left[e^{-i q r_{0,\alpha}} {\bf F}({\bf k}+q\hat{\bf e}_\alpha, {\bf k}) -(q\leftrightarrow -q)\right],
\end{equation}
where $\alpha=x,y,z$.
When expanding the exponential, there is a term that depends on ${\bf r}_0$,
\begin{equation}
\label{eq:zortzi}
\frac{1}{2}\sum_{\bf k} \sum_\alpha ({\bf r}_0\cdot\hat{\bf e}_\alpha) \hat{\bf e}_\alpha \times {\bf F}({\bf k},{\bf k}).
\end{equation}
In a centrosymmetric crystal (like \bise\, or \bite), if we measure the position from the inversion center, for every nucleus at ${\bf r}_0$ there is another identical nucleus at $-{\bf r}_0$. 
Equation~(\ref{eq:zortzi}) reverses sign under ${\bf r}_0\to-{\bf r}_0$ and, since the measured NMR signal averages over all identical nuclei, Eq.~(\ref{eq:zortzi}) is averaged out.
In a noncentrosymmetric crystal, Eq.~(\ref{eq:zortzi}) is not fully averaged-out, but the average remains finite. 

The expression for $\delta B^{\rm dip}_x({\bf r}_0)$ is identical to Eq.~(\ref{eq:dibost}), provided that we replace  $M^z_{\kv,\kv'}(j_0)$ by
\begin{align}
\label{eq:disei}
M^x_{\kv,\kv'}(j_0)&=
\sum_{\mu \mu'} \sum_{\sigma \sigma'} \parS{
\beta(j_0)(-2\delta_{\mu p_x}\delta_{\mu' p_x}+\delta_{\mu p_y}\delta_{\mu' p_y}+\delta_{\mu p_z}\delta_{\mu' p_z})\langle\sigma|\sigma^x|\sigma'\rangle
-3\beta' (j_0)(\delta_{\mu p_x}\delta_{\mu' p_y}+\delta_{\mu p_y}\delta_{\mu' p_x})\langle\sigma|\sigma^y|\sigma'\rangle
\right.\nonumber\\ 
&\left. \qquad\qquad
-3\beta' (j_0)(\delta_{\mu p_x}\delta_{\mu' p_z}+\delta_{\mu p_z}\delta_{\mu' p_x})\langle\sigma|\sigma^z|\sigma'\rangle
}
|\psi_{\kv j_0 \mu \sigma} \rangle \langle \psi_{\kv' j_0 \mu' \sigma'} |
.
\end{align}
Together, Eqs.~(\ref{eq:dibost}), \eqref{eq:mz-dip} and (\ref{eq:disei}) allow a numerical calculation of $\delta{\bf B}^{\rm dip}({\bf r}_0)$.
To do so, it is convenient to follow Appendix~\ref{app:derivationPT} and to insert the relation $\sum_{n''} |u_{{\bf k} n''}\rangle\langle u_{{\bf k} n''}| = {\bf 1}$ in the first line of $\delta B^{\rm dip}_x$ and $\delta B^{\rm dip}_z$.

\section{Details concerning the calculation of the orbital shift} \label{app:orbital} 

In this appendix we present some details of the calculation of the orbital Knight shift, which are important in order to obtain Fig.~\ref{fig:korb}.
We discuss the paramagnetic and the diamagnetic parts separately.

\subsection{Paramagnetic term}

The contribution from the paramagnetic current to the orbital shift is given by
\begin{equation}
\label{eq:bat}
\delta {\bf B}^{\rm par} ({\bf r}_0) = \frac{\mu_0}{4\pi} \int \drm^3 r \langle {\bf J}_P({\bf r})\rangle \times \frac{{\bf r}_0-{\bf r}}{|{\bf r}_0-{\bf r}|^3},
\end{equation}
where ${\bf J}_P=(-e/2)\{{\bf v},|{\bf r}\rangle\langle{\bf r}|\}$ is the paramagnetic current operator and the integral is over entire volume of the crystal.
To first order in the applied field, the expectation value of the paramagnetic current is
\begin{equation}
\langle {\bf J}_P({\bf r})\rangle = \sum_{{\bf k}{\bf k}'} \sum_{ n n'} \langle\psi_{{\bf k} n} |{\bf J}_P ({\bf r})|\psi_{{\bf k}'n'}\rangle\langle\psi_{{\bf k}'n'}|\delta{\cal H}|\psi_{{\bf k} n}\rangle f_{n n'}({\bf k}, {\bf k}'),
\end{equation}
where $\delta{\cal H}= \mu_B {\boldsymbol\sigma}\cdot{\bf B}+(\mu_B m/\hbar) {\bf r}\times{\bf v}$, and $f_{n n'}({\bf k},{\bf k}')$ has been defined in Eq.~(\ref{eq:fnn}).
From Eq.~\eqref{eq:oo} of Appendix~\ref{app:derivationPT}, we have
\begin{align}
\label{eq:batb}
\begin{split}
\langle J_{P, i}({\bf r})\rangle = 
&\mu_B\sum_{{\bf k} n n'}\langle\psi_{{\bf k} n}|J_{P,i}({\bf r})|\psi_{{\bf k} n'}\rangle f_{n n'}({\bf k}, {\bf k}) 
\langle\psi_{{\bf k} n'} |{\boldsymbol\sigma}\cdot{\bf B}|\psi_{{\bf k} n}\rangle\\
+&\frac{m\mu_B}{\hbar^2}{\rm Im}\sum_{{\bf k} n n'}\langle\psi_{{\bf k} n}|J_{P,i}({\bf r})|\psi_{{\bf k} n'}\rangle f_{n n'}({\bf k}, {\bf k}) 
(\langle\grad_{\bf k} u_{{\bf k} n'} |\times(\grad_{\bf k}{\cal H}_{\bf k})|u_{{\bf k} n}\rangle_{\rm cell})\cdot{\bf B}\\
-&\frac{m\mu_B}{\hbar}{\rm Im}\sum_{{\bf k}{\bf k}' n n'}\delta_{{\bf k}{\bf k}'}\left[\grad_{\bf k}\times \left(\langle\psi_{{\bf k} n}|J_{P,i}({\bf r})|\psi_{{\bf k}' n'}\rangle f_{n n'}({\bf k}, {\bf k}') \langle\psi_{{\bf k} n'}|{\bf v}|\psi_{{\bf k} n}\rangle\right)\right]\cdot{\bf B},
\end{split}
\end{align}
for $i=x,y,z$.
Here, we have kept the term $\mu_B{\boldsymbol\sigma}\cdot{\bf B}$ because it contributes to $\langle{\bf J}_P({\bf r})\rangle$ in presence of spin-orbit interaction.

The matrix elements of the paramagnetic current operator are
\begin{align}
\begin{split}
&\langle\psi_{{\bf k} n} |{\bf J}_P({\bf r})|\psi_{{\bf k}' n'}\rangle \simeq \frac{i e \hbar}{2 m}\left[\langle\psi_{{\bf k} n}|{\bf r}\rangle \grad \langle{\bf r}|\psi_{{\bf k}' n'}\rangle - (\grad \langle\psi_{{\bf k} n}|{\bf r}\rangle)\langle{\bf r}|\psi_{{\bf k}' n'}\rangle \right]\\
&=\frac{i e \hbar}{2 m N} \sum_{{\bf R} j \mu \sigma} \sum_{{\bf R}' j' \mu' \sigma'} \langle\psi_{{\bf k} n}|\psi_{{\bf k} j \mu \sigma}\rangle\langle\psi_{{\bf k}' j' \mu' \sigma'}|\psi_{{\bf k}' n'}\rangle e^{i {\bf k}'\cdot({\bf R}'+{\bf t}_{j'})} e^{-i {\bf k}\cdot({\bf R}+{\bf t}_j)}\\
&~~~~~~~~~~~~~~~~~~~~~~~~~~~\left[\langle{\bf R} j \mu \sigma|{\bf r}\rangle \grad\langle{\bf r}|{\bf R}' j' \mu'\sigma'\rangle -  (\grad  \langle{\bf R} j \mu \sigma|{\bf r}\rangle) \langle{\bf r}|{\bf R}'j'\mu'\sigma'\rangle\right].
\end{split}
\end{align}
In the first line, we have approximated $\langle{\bf r}|{\bf v}|\psi_{{\bf k} n}\rangle \simeq -(i\hbar/m)\grad\langle{\bf r}|\psi_{{\bf k} n}\rangle$.
Because $|{\bf R} j \mu\sigma\rangle = |{\bf R}  j \mu\rangle|\sigma\rangle$, we can write $ (\grad \langle{\bf R} j \mu \sigma|{\bf r}\rangle)\langle{\bf r}|{\bf R}'j'\mu'\sigma'\rangle=\delta_{\sigma\sigma'} \langle{\bf r}|{\bf R}'j'\mu'\rangle \grad \langle{\bf R} j \mu |{\bf r}\rangle$. 
In addition, the atomic wave functions are real and exponentially localized.
Therefore, we can approximate
\begin{equation}
\label{eq:bi}
\langle\psi_{{\bf k} n} |{\bf J}_P({\bf r})|\psi_{{\bf k}' n'}\rangle \simeq \frac{i e \hbar}{2 m N}\sum_{{\bf R} j \mu\mu' \sigma} e^{i ({\bf k'-k})\cdot({\bf R}+{\bf t}_j)}
\langle\psi_{{\bf k} n}|\psi_{{\bf k} j \mu \sigma}\rangle\langle\psi_{{\bf k}' j \mu' \sigma}|\psi_{{\bf k}' n'}\rangle
\left[\langle {\bf r}|{\bf R} j \mu\rangle \grad\langle{\bf r}|{\bf R} j \mu'\rangle - (\mu\leftrightarrow \mu')\right].
\end{equation}
Next, we assume that the main contribution to the total orbital shift at ${\bf r}_0$ comes from the current density in the vicinity of ${\bf r}_0$. 
This allows us to estimate Eq.~(\ref{eq:bat}) via
\begin{equation}
\langle\psi_{{\bf k} n} |{\bf J}_P({\bf r}\simeq {\bf r}_0)|\psi_{{\bf k}' n'}\rangle \simeq \frac{i e \hbar}{2 m N}  e^{i ({\bf k'-k})\cdot{\bf r}_0} \sum_{\mu\mu' \sigma}\langle\psi_{{\bf k} n}|\psi_{{\bf k} j_0 \mu \sigma}\rangle\langle\psi_{{\bf k}' j_0 \mu' \sigma}|\psi_{{\bf k}' n'}\rangle
\left[\langle {\bf r}|{\bf R}_0 j_0 \mu\rangle \grad\langle{\bf r}|{\bf R}_0 j_0 \mu'\rangle - (\mu\leftrightarrow \mu')\right].
\end{equation}
Hence, the spatial integral that we need to compute is 
\begin{align}
\label{eq:bib}
&\int \drm^3 r\frac{1}{|{\bf r}_0-{\bf r}|^3} \langle\psi_{{\bf k} n}|{\bf J}_P({\bf r})|\psi_{{\bf k}' n'}\rangle \times({\bf r}_0-{\bf r})\nonumber\\
&\simeq \frac{i e\hbar}{2 m N} e^{i({\bf k}'-{\bf k})\cdot{\bf r}_0} \sum_{\mu\mu'\sigma} \langle\psi_{{\bf k} n}|\psi_{{\bf k} j_0 \mu\sigma}\rangle\langle\psi_{{\bf k}' j_0\mu'\sigma}|\psi_{{\bf k}'n'}\rangle
\int \drm^3 r  \frac{\left[\langle {\bf r}|{\bf R}_0 j_0 \mu\rangle \grad\langle{\bf r}|{\bf R}_0 j_0 \mu'\rangle - (\mu\leftrightarrow \mu')\right]\times({\bf r}_0-{\bf r})}{|{\bf r}_0-{\bf r}|^3}\nonumber\\
&=-\frac{i e\hbar}{2 m N} e^{i({\bf k}'-{\bf k})\cdot{\bf r}_0} \sum_{\mu\mu'\sigma} \langle\psi_{{\bf k} n}|\psi_{{\bf k} j_0 \mu\sigma}\rangle\langle\psi_{{\bf k}' j_0\mu'\sigma}|\psi_{{\bf k}'n'}\rangle
\int \drm^3 r \frac{\left[\langle {\bf r}|{\bf 0} j_0 \mu\rangle \grad\langle{\bf r}|{\bf 0} j_0 \mu'\rangle - (\mu\leftrightarrow \mu')\right]\times{\bf r}}{r^3},
\end{align}
where we have exploited the fact that $\langle{\bf r}|{\bf R}_0 j_0 \mu\rangle$ is a function of ${\bf r}-{\bf R}_0-{\bf t}_{j_0}={\bf r}-{\bf r}_0$.
By definition, $\langle{\bf r}|{\bf 0} j_0 \mu\rangle$ is the wave function corresponding to the orbital $\mu$ of atom $j_0$, the atomic center being at the origin of the coordinate system.
Next, let us concentrate on 
\begin{equation}
\label{eq:ole}
\int \drm^3 r \frac{1}{r^3} \left[\langle {\bf r}|{\bf 0} j_0 \mu\rangle \grad\langle{\bf r}|{\bf 0} j_0 \mu'\rangle - (\mu\leftrightarrow \mu')\right]\times{\bf r},
\end{equation}
one component at a time.
The $z$ component reads
\begin{equation}
\int \drm^3 r \frac{1}{r^3} \left[\langle {\bf r}|{\bf 0} j_0 \mu\rangle \partial_x\langle{\bf r}|{\bf 0} j_0 \mu'\rangle y -\langle {\bf r}|{\bf 0} j_0 \mu\rangle \partial_y\langle{\bf r}|{\bf 0} j_0 \mu'\rangle x - (\mu\leftrightarrow \mu')\right].
\end{equation}
The angular integration leads to the following selection rule: $\mu=p_x$ and $\mu'=p_y$, OR $\mu=p_y$ and $\mu'=p_x$.
We can factorize $\langle{\bf r}|{\bf 0} j_0 p_x\rangle = R_{p, j_0}(r) x/r$ and  $\langle{\bf r}|{\bf 0} j_0 p_y\rangle = R_{p, j_0}(r) y/r$, where $R_{p, j_0}$ is the radial part of the wave function (depending only on $|{\bf r}|$).
Then, some simple algebra leads to
\begin{align}
\label{eq:hiru}
\begin{split}
&\int \drm^3 r \frac{1}{r^3} \left[\langle {\bf r}|{\bf 0} j_0 \mu\rangle \partial_x\langle{\bf r}|{\bf 0} j_0 \mu'\rangle y -\langle {\bf r}|{\bf 0} j_0 \mu\rangle \partial_y\langle{\bf r}|{\bf 0} j_0 \mu'\rangle x - (\mu\leftrightarrow \mu')\right]\\
&=-2(\delta_{\mu p_x}\delta_{\mu'p_y}-\delta_{\mu p_y}\delta_{\mu' p_x})\int \drm^3 r R^2_{p, j_0}(r)\frac{x^2}{r^5},
\end{split}
\end{align}
where we have used $\int \drm^3 r R_{p, j_0}(r)\, x^2/r^5=\int \drm^3 r R_{p, j_0}(r)\, y^2/r^5$. 
Likewise, the $x$ component of Eq.~(\ref{eq:ole}) reads
\begin{align}
\label{eq:lau}
\begin{split}
&\int \drm^3 r \frac{1}{r^3} \left[\langle {\bf r}|{\bf 0} j_0 \mu\rangle \partial_y\langle{\bf r}|{\bf 0} j_0 \mu'\rangle z -\langle {\bf r}|{\bf 0} j_0 \mu\rangle \partial_z\langle{\bf r}|{\bf 0} j_0 \mu'\rangle y - (\mu\leftrightarrow \mu')\right]\\
&=-2(\delta_{\mu p_y}\delta_{\mu' p_z}-\delta_{\mu p_z}\delta_{\mu' p_y})\int \drm^3 r R^2_{p, j_0}(r)\frac{x^2}{r^5}.
\end{split}
\end{align}
Combining Eqs.~(\ref{eq:bat}), (\ref{eq:batb}),  (\ref{eq:bib}), (\ref{eq:hiru}) and (\ref{eq:lau}), we obtain
	\begin{align}
	\label{eq:bost}
\begin{split}
&\delta B^{\rm par}_z ({\bf r}_0)
\simeq
\frac{i}{V}\sum_{{\bf k} n n'}
\langle\psi_{{\bf k} n}| \tilde{M}^z_{ {\bf k},{\bf k}} (j_0) |\psi_{{\bf k} n}\rangle
f_{n n'}({\bf k}, {\bf k})\langle\psi_{{\bf k} n'}|{\boldsymbol\sigma}\cdot{\bf B}|\psi_{{\bf k} n}\rangle
\\
&+\frac{m}{\hbar^2}  {\rm Re}\frac{1}{V}\sum_{{\bf k} n n'}
\langle\psi_{{\bf k} n}|\tilde{M}^z_{ {\bf k},{\bf k}} (j_0)|\psi_{{\bf k} n'}\rangle
f_{n n'}({\bf k}, {\bf k})\langle\grad_{\bf k} u_{{\bf k} n'}|\times(\grad_{\bf k} {\cal H}_{\bf k})|u_{{\bf k} n}\rangle_{\rm cell}\cdot{\bf B}
\\
&-\frac{m}{\hbar^2} {\rm Re}\frac{1}{V} \sum_{{\bf k}{\bf k}' n n'}\delta_{{\bf k}{\bf k}'}\big\{\grad_{\bf k}\times\left[\langle\psi_{{\bf k} n}|\tilde{M}^z_{ {\bf k},{\bf k}'} (j_0)|\psi_{{\bf k}' n}\rangle f_{n n'}({\bf k}, {\bf k}') \langle\psi_{{\bf k} n'}|\hbar{\bf v}|\psi_{{\bf k} n}\rangle\right]\big\}\cdot{\bf B},
\end{split}
\end{align}
where we have defined 
\begin{equation}
\tilde{M}^z_{\kv, \kv'}(j_0)=\gamma(j_0) \sum_{\mu,\mu'}\sum_{\sigma}(\delta_{\mu p_x} \delta_{\mu' p_y}-\delta_{\mu p_y}\delta_{\mu' p_x}) |\psi_{\kv j_0 \mu \sigma} \rangle \langle \psi_{\kv j_0 \mu' \sigma} |
\end{equation}
 and 
\begin{equation}
\label{eq:zazpi}
\gamma (j_0)\equiv V_{\rm cell}\frac{\mu_0 \mu_B^2}{2\pi}\int \drm^3 r R^2_{j_0,p}(r)\frac{x^2}{r^5} = 5 \beta(j_0).
\end{equation}
In the derivation of Eq.~(\ref{eq:bost}), we have used ${\rm Im}(i z)= {\rm Re} (z)$, $\mu_B = e \hbar/(2 m)$ and  $N=V/V_{\rm cell}$, where $V_{\rm cell}$ is the unit cell volume.
In addition, we have set $\exp[i({\bf k}'-{\bf k})\cdot{\bf r}_0] \to 1$ (cf. Appendix~\ref{app:dipolar} for a justification of this).
The numerical values for $\gamma (j_0)$ may be read off from Eq.~(\ref{eq:betanum}).
Note that $\gamma({\rm Bi})$ takes slightly different values in \bise\, and \bite, because their unit cell volumes are not identical.
The expression for $\delta B^{\rm par}_x({\bf r}_0)$ is identical to Eq.~(\ref{eq:bost}), provided that we replace $\tilde{M}^z_{\mu\sigma,\mu'\sigma'}$ by
\begin{equation}
\tilde{M}^x_{\kv,\kv'}(j_0)= \gamma(j_0) \sum_{\mu \mu'}\sum_\sigma(\delta_{\mu p_y} \delta_{\mu' p_z}-\delta_{\mu p_z}\delta_{\mu' p_y})
|\psi_{\kv j_0 \mu \sigma} \rangle \langle \psi_{\kv j_0 \mu' \sigma} |
.
\end{equation}
Together, Eqs.~(\ref{eq:bost}) and (\ref{eq:zazpi}) allow a numerical calculation of $\delta{\bf B}^{\rm par}({\bf r}_0)$.
To do so, it is convenient to follow Appendix~\ref{app:derivationPT} and to insert the relation $\sum_{n''} |u_{{\bf k} n''}\rangle\langle u_{{\bf k} n''}| = {\bf 1}$ in the first line of $\delta B^{\rm par}_x$ and $\delta B^{\rm par}_z$.

\subsection{Diamagnetic term}\label{subsec:dia}

The contribution from the paramagnetic current to the orbital shift is given by
\begin{equation}
\label{eq:batc}
\delta {\bf B}^{\rm dia} ({\bf r}_0) = \frac{\mu_0}{4\pi} \int \drm^3 r \langle {\bf J}_D({\bf r})\rangle \times \frac{{\bf r}_0-{\bf r}}{|{\bf r}_0-{\bf r}|^3},
\end{equation}
where ${\bf J}_D ({\bf r})=-(e^2/m) {\bf A}({\bf r})|{\bf r}\rangle\langle{\bf r}|$ is the diamagnetic current operator and the integral is over entire volume of the crystal.
To first order in the applied field, the expectation value of the diamagnetic current is
\begin{align}
\langle{\bf J}_D({\bf r})\rangle=&-\frac{e^2}{m} {\bf A}({\bf r})\sum_{{\bf k} n} f_{{\bf k} n} \langle\psi_{{\bf k} n}|{\bf r}\rangle\langle{\bf r}|\psi_{{\bf k} n}\rangle\nonumber\\
=&-\frac{e^2}{m} {\bf A}({\bf r})\sum_{{\bf k} n } f_{{\bf k} n}\sum_{j\mu\sigma}\sum_{j'\mu'}\langle\psi_{{\bf k} n}|\psi_{{\bf k} j \mu\sigma}\rangle\langle\psi_{{\bf k} j'\mu'\sigma}|\psi_{{\bf k} n}\rangle
\frac{1}{N}\sum_{{\bf R}{\bf R}'} e^{i{\bf k}\cdot({\bf R}'+{\bf t}_{j'}-{\bf R}-{\bf t}_j)} \langle{\bf R} j \mu|{\bf r}\rangle\langle{\bf r}|{\bf R}' j' \mu'\rangle,
\end{align}
where in the second line we have used the expansion of the states $|\psi_{\kv n}\rangle$ in the localized atomic orbital basis, as well as $\langle{\bf R} j \mu \sigma|{\bf r}\rangle\langle{\bf r}|{\bf R}' j' \mu' \sigma'\rangle = \langle{\bf R} j \mu |{\bf r}\rangle\langle{\bf r}|{\bf R}' j' \mu' \rangle \delta_{\sigma \sigma'}$.
Given that the atomic wave functions are localized, we approximate
\begin{equation}
\langle{\bf J}_D({\bf r})\rangle\simeq-\frac{e^2}{m} {\bf A}({\bf r})\sum_{{\bf k} n } f_{{\bf k} n}\sum_{j\mu\mu'\sigma}\langle\psi_{{\bf k} n}|\psi_{{\bf k} j \mu\sigma}\rangle\langle\psi_{{\bf k} j\mu'\sigma}|\psi_{{\bf k} n}\rangle
\frac{1}{N}\sum_{{\bf R}}\langle{\bf R} j \mu|{\bf r}\rangle\langle{\bf r}|{\bf R} j \mu'\rangle.
\end{equation}
Once again, we assume that the main contribution to the total orbital shift at ${\bf r}_0$ comes from the current density in the vicinity of ${\bf r}_0$. 
This allows us to estimate Eq.~(\ref{eq:batc}) via
\begin{equation}
\langle{\bf J}_D({\bf r})\rangle\simeq-\frac{e^2}{m} {\bf A}({\bf r})\sum_{{\bf k} n } f_{{\bf k} n}\sum_{\mu\mu'\sigma}\langle\psi_{{\bf k} n}|\psi_{{\bf k} j_0 \mu\sigma}\rangle\langle\psi_{{\bf k} j_0\mu'\sigma}|\psi_{{\bf k} n}\rangle
\frac{1}{N}\langle{\bf R}_0 j_0 \mu|{\bf r}\rangle\langle{\bf r}|{\bf R}_0 j_0 \mu'\rangle.
\end{equation}
Then, 
\begin{equation}
\delta{\bf B}^{\rm dia} ({\bf r}_0)\simeq-\frac{\mu_0 e^2}{8\pi m N}\sum_{{\bf k} n} f_{{\bf k} n} \sum_{\mu\mu'\sigma} \langle\psi_{{\bf k} n}|\psi_{{\bf k} j_0 \mu\sigma}\rangle\langle\psi_{{\bf k} j_0 \mu' \sigma}|\psi_{{\bf k} n}\rangle \int \drm^3 r \frac{1}{|{\bf r}_0-{\bf r}|^3}\left[({\bf B}\times{\bf r})\times({\bf r}_0-{\bf r})\right] \langle{\bf R}_0 j_0 \mu|{\bf r}\rangle\langle{\bf r}|{\bf R}_0 j_0\mu'\rangle.
\end{equation}
Writing 
$({\bf B}\times{\bf r})\times({\bf r}_0-{\bf r})= ({\bf B}\times({\bf r}-{\bf r}_0+{\bf r}_0))\times({\bf r}_0-{\bf r})$,
and recalling that $\langle{\bf R}_0 j_0 \mu|{\bf r}\rangle\langle{\bf r}|{\bf R}_0 j_0\mu'\rangle$ depends only on $|{\bf r}-{\bf r}_0|$, we have
\begin{align}
\delta{\bf B}^{\rm dia} ({\bf r}_0)
&\simeq
\frac{\mu_0 e^2}{8\pi m N}\sum_{{\bf k} n} f_{{\bf k} n} \sum_{\mu\mu'\sigma} \langle\psi_{{\bf k} n}|\psi_{{\bf k} j_0 \mu\sigma}\rangle\langle\psi_{{\bf k} j_0 \mu' \sigma}|\psi_{{\bf k} n}\rangle \int \drm^3 r \frac{1}{r^3}\left[({\bf B}\cdot{\bf r}){\bf r}-{\bf B} r^2\right] \langle{\bf 0} j_0 \mu|{\bf r}\rangle\langle{\bf r}|{\bf 0} j_0\mu'\rangle,
\end{align}
where we have used ${\bf B}\times{\bf r})\times{\bf r} =({\bf B}\cdot{\bf r}){\bf r}-{\bf B} r^2 $.
Here, we have ignored a term that is linear in ${\bf r}_0$, by virtue of the discussion following Eq.~(\ref{eq:zortzi}).

Then, the relevant spatial integral  is
\begin{equation}
\label{eq:bede}
\int \drm^3 r \frac{1}{r^3}\left[({\bf B}\cdot{\bf r}){\bf r}-{\bf B} r^2\right]\langle{\bf 0} j_0 \mu|{\bf r}\rangle\langle{\bf r}|{\bf 0} j_0\mu'\rangle.
\end{equation}
This integral yields the selection rules for the orbitals.
For instance, the $z$ component of Eq.~(\ref{eq:bede}) is 
\begin{equation}
\label{eq:bedez}
\int \drm^3 r \frac{1}{r^3} [B_x x z + B_y y z - B_z (x^2+y^2)] \langle{\bf 0} j_0 \mu|{\bf r}\rangle\langle{\bf r}|{\bf 0} j_0\mu'\rangle.
\end{equation}
Accordingly, 
an explicit evaluation of the angular integrals yields
\begin{align}
\begin{split}
	\delta B^{\rm dia}_z({\bf r}_0) &\simeq
 	\lambda_{pp}' (j_0)\frac{1}{V}\sum_{{\bf k} n} f_{{\bf k} n} \sum_\sigma 
 	\left[
 		\langle\psi_{{\bf k} n}|\psi_{{\bf k} j_0 p_x \sigma}\rangle\langle\psi_{{\bf k} j_0 p_z \sigma}|\psi_{{\bf k} n}\rangle +(p_x\leftrightarrow p_z)
 	\right] B_x\\
	&\quad
	+\lambda_{pp}' (j_0) \frac{1}{V} \sum_{{\bf k} n} f_{{\bf k} n} \sum_\sigma \left[
		\langle\psi_{{\bf k} n}|\psi_{{\bf k} j_0 p_y \sigma}\rangle\langle\psi_{{\bf k} j_0 p_z \sigma}|\psi_{{\bf k} n}\rangle +(p_y\leftrightarrow p_z)
	\right] B_y\\
	&\quad
	-\lambda_{ss} (j_0) \frac{1}{V}\sum_{{\bf k} n} f_{{\bf k} n} \sum_\sigma |\langle\psi_{{\bf k} n}|\psi_{{\bf k} j_0 s \sigma}\rangle|^2 B_z\\
	&\quad
	-\lambda_{pp} (j_0) \frac{1}{V}\sum_{{\bf k} n} f_{{\bf k} n} \sum_\sigma \left[|\langle\psi_{{\bf k} n}|\psi_{{\bf k} j_0 p_x \sigma}\rangle|^2 + |\langle\psi_{{\bf k} n}|\psi_{{\bf k} j_0 p_y \sigma}\rangle|^2+ \frac{1}{2}|\langle\psi_{{\bf k} n}|\psi_{{\bf k} j_0 p_z \sigma}\rangle|^2\right] B_z,
\end{split}
\end{align}
where the angular integrals can be calculated analytically and the radial integrals estimated numerically leading to the numerical factors
\begin{equation}
\lambda_{ss}(j_0)=V_{\rm cell}\frac{\mu_0 e^2}{8\pi m}\int \drm^3 r R_{s,j_0}^2(r)\frac{x^2+y^2}{r^3}=\left\{\begin{array}{cc}  
0.0043 \text{\AA}^3 & \text{ for $j_0$=Bi in \bise}\\
0.0051\, \text{\AA}^3 & \text{ for $j_0$=Bi in \bite}\\
0.0057\, \text{\AA}^3 & \text{ for $j_0$=Se in \bise}\\
0.0058\, \text{\AA}^3 & \text{ for $j_0$=Te in \bite}
\end{array}
\right.
\end{equation}
\begin{equation}
\lambda_{pp}(j_0)=V_{\rm cell}\frac{\mu_0 e^2}{8\pi m}\int \drm^3 r R^2_{p,j_0}(r)\frac{(x^2+y^2)x^2}{r^5}=
\left\{\begin{array}{cc}  
0.0013 \text{\AA}^3 & \text{ for $j_0$=Bi in \bise}\\
0.0015\, \text{\AA}^3 & \text{ for $j_0$=Bi in \bite}\\
0.0017\, \text{\AA}^3 & \text{ for $j_0$=Se in \bise}\\
0.0018\, \text{\AA}^3 & \text{ for $j_0$=Te in \bite}
\end{array}
\right.
\end{equation}
\begin{equation}
\lambda_{pp}'(j_0)=V_{\rm cell}\frac{\mu_0 e^2}{8\pi m}\int \drm^3 r R_{p,j_0}^2(r)\frac{x^2 z^2}{r^5}=\frac{1}{4}\lambda_{pp}(j_0).
\end{equation}
	Similarly, $\delta B^{\rm dia}_x({\bf r}_0)$ is obtained under the exchange $B_x \leftrightarrow B_z$ and $p_x \leftrightarrow p_z$.
\\

\end{widetext}


\begin{thebibliography}{100}
\bibitem{pert} For simplicity, we sketch the derivation using non-degenerate perturbation theory. However, Eq.~(\ref{eq:tr2}) is applicable for systems with degenerate bands.
\end{thebibliography}
\end{document}